\documentclass[aps,prl,amsmath,amssymb,superscriptaddress,reprint]{revtex4-2}

\usepackage{cmap}
\usepackage[T1]{fontenc}
\usepackage{newtxtext}
\usepackage[varvw]{newtxmath}
\usepackage{microtype}
\usepackage{graphicx}
\graphicspath{ {figures} }
\usepackage{bm}
\usepackage{hyperref}
\hypersetup{colorlinks=true,allcolors=blue}
\usepackage{mathtools}
\usepackage{array}
\usepackage{upgreek}
\usepackage{orcidlink}
\usepackage{makecell}

\renewcommand{\leq}{\leqslant}
\renewcommand{\geq}{\geqslant}

\usepackage{color}
\usepackage{physics, ulem}
\usepackage{cases}
\usepackage{mathrsfs}
\usepackage{subfigure}
\usepackage{comment}
\usepackage{physics2}
\usephysicsmodule{ab}

\newcommand{\lb}{\left(}
\newcommand{\rb}{\right)}

\begin{document}

\title{Nodal magnon spin current at an antiferromagnet/altermagnet junction}

\author{Koki Mizuno\,\orcidlink{0009-0004-8276-1045}}
\email{mizuno.koki.t8@s.mail.nagoya-u.ac.jp}
\affiliation{
    Department of Physics,
    \href{https://ror.org/04chrp450}{Nagoya University},
    Nagoya, Japan
}

\begin{abstract}
    We investigate the thermally driven magnon spin current across a junction between a $d$-wave altermagnet (AM) and a ferromagnet (FM) or an antiferromagnet (AFM).
    The current consists of a component independent of the interface angle and an angularly dependent component with the $d$-wave form.
    For a sublattice-symmetric interface, the two junctions behave oppositely under reversal of the altermagnetic domain: in the FM/AM junction the angularly dependent component is odd and cancels on averaging over the domains, whereas in the AFM/AM junction the angularly independent component vanishes by symmetry and the angularly dependent component survives the domain average at a fixed N\'eel vector of the AFM.
    Both follow from a selective tunneling: a pair of spins bonded across the interface transmits magnons when the spins are parallel and not when antiparallel, which makes the AFM/AM junction a magnon valve: the configuration in which the bonded spins are antiparallel blocks the current up to the sublattice admixture of the magnons.
    We reveal the uniqueness of the altermagnet by comparing with the anisotropic antiferromagnet.
    The anisotropic tunneling is detectable as the sign reversal of the current between the interfaces normal to $[310]$ and $[130]$, and the valve reads the N\'eel vector of the altermagnet relative to that of the antiferromagnet through the magnitude of the current at a fixed interface.
\end{abstract}

\maketitle

\textit{Introduction.}---
The altermagnet (AM), a compensated antiferromagnet with anisotropic spin splitting without spin-orbit coupling \cite{Altermagnet_2022__mejkal_1,Altermagnet_2022__mejkal_2,Altermagnet_material_2024_Tomas,Altermagnet_2024_Bai, Cite-in-Almag-S-E-conv_2019_Naka}, is one of the leading candidates for next-generation magnetic memory devices capable of high-density storage because of the absence of a magnetic field leakage \cite{Magnon-tunneling_ref-in-paper_memory_2025_Chen,Magnon-tunneling_ref-in-paper_memory_2025_Guo,Magnon-tunneling_ref-in-paper_memory_2026_Jungwirth}.
Since the technology of information reading is essential for memory, in this paper, we examine the magnon spin current as a medium for carrying information.
The electron transport in AM is attracting much attention as a method of information reading \cite{Magnon-tunneling_ref-in-paper_AFM-spintronics-domain-readout_2024_Betancourt, Magnon-tunneling_ref-in-paper_AFM-spintronics-domain-readout_2022__mejkal, Magnon-tunneling_ref-in-paper_AFM-spintronics-domain-readout_2018_Baltz}, but the Joule heating is unavoidable.
On the other hand, the magnon tunneling between a ferromagnet (FM) or an antiferromagnet (AFM) and an AM, the focus of this letter, is applicable even to information transmission in insulating systems and would avoid the Joule heating.
Coherent transport between magnon condensates has been formulated as a Josephson effect \cite{Magnon-tunneling_ref-in-paper_magnon-Josephson_2014_Nakata, Magnon-tunneling_ref-in-paper_magnon-Josephson_2021_Kreil}.
The incoherent thermal magnon current across an FM/AM or AFM/AM interface has not.

The AM was conceived as the magnetic counterpart of an unconventional superconductor (SC) \cite{Altermagnet_2022__mejkal_1,Altermagnet_2022__mejkal_2}: a compensated collinear magnet whose bands have a $d$-wave (or higher even-parity) spin splitting.
The same even-parity anisotropy appears also in the magnon spectrum \cite{Magnon-tunneling_ref-in-paper_AM-thermal-magnon-transport_2025_Galindez-Ruales, Smejkal_2023_PhysRevLett, McClarty_2025_PhysRevB}.
For unconventional ($d$-wave) SC, bulk thermodynamics established the presence of nodes but not the sign structure of the order parameter.
The decisive evidence came from junctions whose transport depends on the interface orientation relative to the nodes \cite{d-wave-SC_1993_Wollman, d-wave-SC_1995_Harlingen, d-wave-SC_1992_Sigrist, d-wave-SC_1996_Kashiwaya}.
Here, we apply this junction-transport argument to the AM \cite{note_sc_analogy}.
In contrast to the SC, the bulk magnon transport in altermagnets is already anisotropic \cite{Magnon-tunneling_ref-in-paper_Seebeck-Nernst_2023_Cui, Magnon-tunneling_ref-in-paper_AM-thermal-magnon-transport_2025_Galindez-Ruales}, but that anisotropy can be averaged out over magnon branches or by averaging over time-reversed domains.

Transport phenomena in junction systems have been pursued for the conduction electrons of metallic AMs \cite{Magnon-tunneling_ref-in-paper_AM-interface-spin-current_2024_Reyes-Osorio,Magnon-tunneling_ref-in-paper_AM-interface-spin-current_2026_Reyes-Osorio}, and the anisotropic spin pumping from an FM insulator into a metallic AM \cite{Altermagnet_spin-current_2023_Sun}.
For thermally driven magnon transport, the bulk magnon Seebeck and Nernst effects \cite{Altermagnet-magnon-tunneling_2024_Wei_enhofer, Altermagnet-magnon-tunneling_2026_Yang, Magnon-tunneling_ref-in-paper_Seebeck-Nernst_2023_Cui, Magnon-tunneling_ref-in-paper_AM-thermal-magnon-transport_2025_Galindez-Ruales} have been reported.
For the magnon transport in junction systems without AMs, the magnon valve effect set by the order configuration of FM or AFM junctions \cite{Magnon-tunneling_ref-in-paper_Magnon-spin-valve_2023_Zhang,Magnon-tunneling_ref-in-paper_Magnon-spin-valve_2021_Yan,Magnon-tunneling_ref-in-paper_Magnon-spin-valve_2018_Guo} has been established.
However, properties of the altermagnetic magnon transport in junction systems with other magnetic orders have not been examined yet.
In this letter, we will answer the two open questions: (i) whether the anisotropy is captured by the interface orientation, and (ii) whether the anisotropy survives the domain average of AM.

First, the interface angle dependence of the magnon spin current exists in both FM/AM and AFM/AM junctions.
However, the FM/AM junction has an interface angle independent contribution, while this contribution vanishes by symmetry in the AFM/AM junction.
Second, after averaging over the magnetic domains of AM with the same weight, the interface angle dependence is cancelled in the FM/AM junction whereas it survives in the AFM/AM junction.
The contrast between the two junction partners is the magnon valve effect by energy conservation on the bonded spins at the interface.
When the bonded spins are parallel, the magnons tunnel through the junction mainly by the particle--particle process.
In contrast, when the spins are antiparallel, this particle--particle path is largely blocked, and the remaining contribution is the sublattice admixture induced by the antiferromagnetic exchange inside the material.
Because that contribution is small, the spin current is suppressed.
Therefore, whereas the valve effect disappears in the FM/AM junction, it exists in the AFM/AM junction depending on whether the N\'eel vector (the order parameter of the antiferromagnetic state) of the AM is parallel or antiparallel to the N\'eel vector of the AFM.

\begin{figure}
    \centering
    \subfigure[FM/AM junction]{\includegraphics[width=0.48\columnwidth]{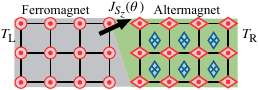}}%
    \hfill
    \subfigure[AFM/AM junction]{\includegraphics[width=0.48\columnwidth]{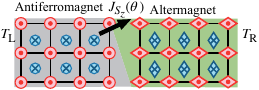}}
    \caption{
        Schematic drawing of the (a) FM/AM and (b) AFM/AM junctions, viewed along the $z$ axis.
        Red symbols with a dot (blue symbols with a cross) denote spins pointing out of (into) the plane; the sublattice B sits at the center of the square formed by the sublattice A, and in the altermagnet the rhombi are elongated along the direction of the stronger intra-sublattice exchange, $x$ on A and $y$ on B.
        The arrow $j_{S_{z}}(\theta)$ denotes the spin current $j_{S_{z}}(\theta)$ through the interface at the interface angle $\theta$.
    }
    \label{fig:system}
\end{figure}

\textit{Magnon tunneling at junctions.}---
Throughout this letter, we use units with $\hbar = k_{\rm B} = 1$.
Here, we consider a junction system composed of two magnetic materials, a ferromagnet (FM) or an antiferromagnet (AFM), and an altermagnet (AM).
The left side is an FM or AFM, and the right side is an AM as shown in Fig.~\ref{fig:system}.
The two materials are connected by a weak exchange interaction at the interface.
The Hamiltonian of the system is given by
\begin{align}
    H = H_{\rm L}^{\rm FM / AFM} + H_{\rm R}^{\rm AM} + H_{\rm I},
\end{align}
where $H_{\rm L}^{\rm FM / AFM}$ and $H_{\rm R}^{\rm AM}$ are the Hamiltonians of the left (FM or AFM) and right (AM) materials, respectively, and $H_{\rm I}$ is the interaction Hamiltonian at the interface.
The left and right materials are described by the Heisenberg model
\begin{align}
    H_{\rm a}^{\rm FM / AFM / AM}
    =&
    -\sum_{n \in \mathbb{Z}} \sum_{l,l' \in \{A,B\}} \sum_{\vec{\delta}}
    J^{\rm a}_{n,\vec{\delta},l,l'}
    \vb{S}^{\rm a}_{\vec{r}_{n,l}} \cdot \vb{S}^{\rm a}_{\vec{r}_{n,l'}+\vec{\delta}}
    \notag\\ 
    &-
    K \sum_{n \in \mathbb{Z}} \sum_{l \in \{A,B\}}
    \left( S^{\rm a}_{z,\vec{r}_{n,l}} \right)^2,
\end{align}
where $J^{\rm a}_{n,\vec{\delta},l,l'}$ is the exchange interaction strength between the spins at sites $\vec{r}_{n,l}$ and $\vec{r}_{n,l'}+\vec{\delta}$ in material $\mathrm{a}$ (L or R), and $\vb{S}^{\rm a}_{\vec{r}_{n,l}}$ is the spin operator at site $\vec{r}_{n,l}$ in material $\mathrm{a}$ satisfying the spin algebra given by $[S_{\mu,\vec{r}_{n,l}}^{\rm a}, S_{\nu,\vec{r}_{m,l'}}^{\rm b}] = i\sum_{\rho} \epsilon_{\mu\nu\rho} S_{\rho,\vec{r}_{n,l}}^{\rm a} \delta_{a,b} \delta_{l,l'} \delta_{n,m}$.
The site $\vec{r}_{n,l}$ is defined as $\vec{r}_{n,l} = n_x\vec{a}_x+n_y\vec{a}_y+\vec{r}_l$, where $n_x,n_y\in\mathbb{Z}$, $\vec{a}_x$ and $\vec{a}_y$ are the primitive vectors of the lattice, and $\vec{r}_l$ is the position of the sublattice $l$ in the unit cell.
For the ferromagnet, the spin-wave Hamiltonian is given by
\begin{align}
    H_{\rm a}^{\rm FM}
    = \sum_{\vec{k}} \mqty( \phi_{\vec{k}, A}^{\dag} & \phi_{\vec{k}, B}^{\dag} )
    \mqty( 
        \gamma_{\rm F}(\vec{k}) & \gamma_{0}(\vec{k}) \\
        \gamma_{0}(\vec{k}) & \gamma_{\rm F}(\vec{k})
    )
    \mqty( \phi_{\vec{k}, A} \\ \phi_{\vec{k}, B} ),
    \label{eq:spin-wave-Hamiltonian-FM}
\end{align}
where $\gamma_{\rm F}(\vec{k}) = (4J_{0} + 2K)S + 2JS \sum_{\mu} (1 - \cos(ak_{\mu}))$ and $\gamma_{0}(\vec{k}) = -4J_{0}S \cos(ak_x/2)\cos(ak_y/2)$.
Here the sublattice A forms a square lattice of constant $a$ with the intra-sublattice exchange $J$ along $x$ and $y$, and the sublattice B is displaced from A by $(a/2, a/2)$, so that the inter-sublattice bonds $J_{0}$ run along the diagonals; in the altermagnet the intra-sublattice exchange is $J \mp \delta J$ along $x$ and $y$ on A and $J \pm \delta J$ on B.
The operator $\phi_{\vec{k}, l}$ is the annihilation operator of the magnon at momentum $\vec{k}$ and sublattice $l$ in the ferromagnet, which satisfies the bosonic commutation relation $[\phi_{\vec{k}, l}, \phi_{\vec{k}', l'}^{\dag}] = \delta_{\vec{k}, \vec{k}'} \delta_{l, l'}$.
The spin-wave Hamiltonian of the antiferromagnet and altermagnet is given by
\begin{align}
    H_{\rm a}^{\rm AFM / AM}
    = \sum_{\vec{k}} \Psi_{\vec{k}}^{\rm a \dag}
    \mqty(
        \gamma_{\rm A}(\vec{k}) & \gamma_{0}(\vec{k}) \\
        \gamma_{0}(\vec{k}) & \gamma_{\rm B}(\vec{k})
    )
    \Psi_{\vec{k}}^{\rm a},
    \quad
    \Psi_{\vec{k}}^{\rm a} = \mqty( \psi^{\rm a}_{\vec{k}, A} \\ \psi^{\rm a}_{\vec{k}, B} ),
    \label{eq:spin-wave-Hamiltonian-AF}
\end{align}
where $\gamma_{\rm A}(\vec{k}) = (-4J_{0} + 2K)S + 2JS \sum_{\mu} (1 - \cos(ak_{\mu})) + 2\delta J S (\cos(ak_x) - \cos(ak_y))$ and $\gamma_{\rm B}(\vec{k}) = (-4J_{0} + 2K)S + 2JS \sum_{\mu} (1 - \cos(ak_{\mu})) - 2\delta J S (\cos(ak_x) - \cos(ak_y))$.
The field $\psi^{\rm a}_{\vec{k}, l} = S^{\rm a}_{+, \vec{k}, l}/\sqrt{2S}$ is the Fourier transform of the spin-raising operator on the sublattice $l$.
On a sublattice whose moment points up (down), it is the annihilation (creation) operator of a canonical Holstein--Primakoff magnon, so that it satisfies $[\psi^{\rm a}_{\vec{k}, l}, \psi^{\rm b \dag}_{\vec{k}', l'}] = \zeta_{\rm a} \sigma_{l} \delta_{\rm a, \rm b} \delta_{\vec{k}, \vec{k}'} \delta_{l, l'}$, where $\sigma_{A} = -\sigma_{B} = 1$ and $\zeta_{\rm a} = \pm 1$ labels the magnetic domain of the material $\mathrm{a}$ by the direction of the moment on the A sublattice.
We fix $\zeta_{\rm L} = +1$ for the antiferromagnet and write $\zeta_{\rm AM}$ for the altermagnet, whose two domains are related by time reversal.
With this choice, the domain reversal keeps the sublattice order, the $2\times 2$ kernel $h^{\rm a}(\vec{k})$ of Eq.~\eqref{eq:spin-wave-Hamiltonian-AF}, and the interface coupling, and enters only through the metric $\zeta_{\rm a} \Sigma_{z}$ of the spinor, $\Sigma_{z} = \mathrm{diag}(\sigma_{A}, \sigma_{B})$, which also determines the retarded Green's function $[(\omega + i\eta) \zeta_{\rm a} \Sigma_{z} - h^{\rm a}(\vec{k})]^{-1}$ of the bulk; the magnon on the sublattice $l$ carries $S_{z} = -\zeta_{\rm a} \sigma_{l}$.
The ferromagnet is the case $\sigma_{A} = \sigma_{B} = 1$ with $\zeta = 1$, for which $\Psi$ reduces to the ordinary bosons $\phi$ of Eq.~\eqref{eq:spin-wave-Hamiltonian-FM}.
Here, $\delta J$ is the sublattice-dependent anisotropy of the same-sublattice exchange, $J^{\rm a}_{n,\vec{\delta},l,l} = J \mp \sigma_{l}\delta J$ for $\vec{\delta} = \pm a\hat{x}$ and $\pm a\hat{y}$, respectively; the antiferromagnet has $\delta J = 0$ and the altermagnet has $\delta J \neq 0$.
The derivation of the spin-wave Hamiltonian and of its discretized form for an arbitrary interface angle $\theta$ is given in the Supplemental Material~\cite{SM}.

The interaction Hamiltonian is given by
\begin{align}
    H_{\mathrm I}
    =
    - \sum_{\langle n,n' \rangle} \sum_{l,l' \in \{A,B\}} \mathcal{J}_{n,n',l,l'}
    \vb{S}^{\rm L}_{\vec{r}_{n,l}} \cdot \vb{S}^{\rm R}_{\vec{r}_{n',l'}},
\end{align}
where a positive $\mathcal{J}_{n,n',l,l'}$ is ferromagnetic, and $\vb{S}^{\rm L}_{\vec{r}_{n,l}}$ and $\vb{S}^{\rm R}_{\vec{r}_{n',l'}}$ are the spin operators at the interface sites $\vec{r}_{n,l}$ and $\vec{r}_{n',l'}$.
As the simplest case, we consider a nearest-neighbor interface coupling, $\mathcal{J}_{n,n',l,l'} = \mathcal{J}_{l}\delta_{l,l'}\delta_{n,n'}$.
This same-sublattice coupling and the cross-sublattice coupling $\mathcal{J}_{\rm AB} = \mathcal{J}_{\rm BA}$ are related by the relabeling $\mathrm{A} \leftrightarrow \mathrm{B}$ of the altermagnet, which reverses $\zeta_{\rm AM}$ and $\delta J$; the two registries give the same currents with the roles of the two domains exchanged. The mapped currents agree exactly at $\theta = \pi/4$, and elsewhere differ by up to $0.24\%$ for the transmitting domain and the FM/AM junction and $1.3\%$ for the blocked domain, the difference being the in-plane displacement of the bonded sites \cite{SM}, and we present the same-sublattice case.
The Holstein--Primakoff hopping used in the current is
$J^{\rm ex}_{l,l'} = -\sqrt{S_{\rm L}S_{\rm R}}\,\mathcal{J}_{l,l'}$.

In this paper, we consider the spin current through the interface carried by magnon tunneling.
The spin current is defined by the time derivative of the total spin in the right material,
\begin{align}
    \hat{j}_{S_{z}}
    &= \sum_{i} \frac{d}{dt} S_{z,i}^{\rm R} 
    = \sum_{\langle i,j \rangle}\frac{\mathcal{J}_{i,j}}{2i}
        \left(
S_{+,i}^{\rm L} S_{-,j}^{\rm R} - S_{-,i}^{\rm L} S_{+,j}^{\rm R}
        \right),
\end{align}
where $S_{\pm,i}^{\rm a} = S_{x,i}^{\rm a} \pm i S_{y,i}^{\rm a}$ is the spin raising/lowering operator.
The expectation value of the spin current is $j_{S_{z}}(\theta) := \langle \hat{j}_{S_{z}} \rangle$, where $\theta$ denotes the interface angle, which defines the interface normal as $\vb*{n} = (\cos\theta, \sin\theta)$ in the $xy$ plane of the altermagnetic anisotropy, i.e. the junction is lateral.
The dependence on $\theta$ is modeled by a continuum along the interface with a lattice discretization along $\vb*{n}$ \cite{SM}, in which every layer contains both sublattices, so that the model describes a compensated interface.
In the lattice, the interfaces normal to $[hk0]$ with even $h + k$ are compensated: they contain both sublattices with equal density, and the in-plane translation that exchanges the sublattices guarantees $\mathcal{J}_{\rm A} = \mathcal{J}_{\rm B}$, whereas those with odd $h + k$, including $[100]$, terminate on a single sublattice.
This expectation value is obtained by the Keldysh Green's function method \cite{Magnon-tunneling_ref-in-paper_Interface-Keldysh-SSE-theory_2018_Matsuo, PhysRevResearch.3.043058} with the interface coupling resummed by the Dyson equation, and reads
\begin{align}
    j_{S_{z}}(\theta)
    &= -\mathrm{Re}\int^{k_{\rm c}}_{-k_{\rm c}} \frac{dk}{2\pi} \int^{\infty}_{-\infty} \frac{d\omega}{2\pi} 
    \mathrm{Tr}\left[
    V G^{\rm K}_{\rm LR}(k,\omega,\theta) 
    \right],
    \label{eq:spin-current}
\end{align}
where $k_{\rm c}$ is the cutoff momentum.
The hopping matrix \(V\) that enters the current is
\begin{align}
    V =
    \begin{pmatrix}
        J^{\rm ex}_{AA}     & 0 \\
        0 & J^{\rm ex}_{BB}
    \end{pmatrix} ,
\end{align}
where $J^{\rm ex}_{\alpha\beta}=-\sqrt{S_{\rm L}S_{\rm R}}\,\mathcal{J}_{\alpha\beta}$ is the Holstein--Primakoff hopping between the sublattices $\alpha$ and $\beta$ at the interface.
The momentum $\vb{k}_{\parallel} = k(\sin\theta, -\cos\theta)$ is the momentum parallel to the interface.
The Keldysh component $G^{\rm K}_{\rm LR}(k,\omega,\theta)$ is the Fourier transform of the Holstein--Primakoff boundary-spinor correlator $G_{\rm LR}(t_1, t_2) = -i\langle T_{\mathcal C} \chi^{\rm L}(t_{1}) \chi^{\rm R\dagger}(t_{2}) \rangle$, where $\chi$ is the boundary Holstein--Primakoff spinor built from $\phi$ or $\psi$ \cite{SM} and $T_{\mathcal C}$ is the time-ordering operator on the Keldysh contour $\mathcal C$. The contraction above is $\mathrm{Tr}_{\rm R}[V_{\rm RL}G^{\rm K}_{\rm LR}]$ with $V_{\rm RL}=V^{\dagger}$; for the diagonal real $V$ used here, $V_{\rm RL}=V$.
This correlation function can be obtained by the Dyson equation with perturbation expansion with respect to the interaction Hamiltonian $H_\mathrm{I}$, of which the details are given in the SM~\cite{SM}.
With $I$ denoting the $2\times 2$ identity, this Keldysh Green's function is given by
\begin{align}
   &G^{\rm K}_{\rm LR}(k,\omega,\theta)
   \notag\\ 
   &=
   \left( I - G_{\rm LR}^{\rm rr} V \right)^{-1} F_{\rm LR}(k,\omega,\theta) \left( I - V G_{\rm LR}^{\rm aa} \right)^{-1},
   \label{eq:Keldysh-Green}
\end{align}
where $G_{\rm LR}^{\rm rr} = g_{\rm L}^{\rm ret}(k,\omega,\theta) V g_{\rm R}^{\rm ret}(k,\omega,\theta)$, $G_{\rm LR}^{\rm aa} = g_{\rm L}^{\rm adv}(k,\omega,\theta) V g_{\rm R}^{\rm adv}(k,\omega,\theta)$, and $F_{\rm LR}(k,\omega,\theta) = g_{\rm L}^{\rm ret}(k,\omega,\theta) V g_{\rm R}^{\rm K}(k,\omega,\theta) + g_{\rm L}^{\rm K}(k,\omega,\theta) V g_{\rm R}^{\rm adv}(k,\omega,\theta)$.
The fluctuation-dissipation theorem gives the Keldysh component of the Green's function in terms of the retarded and advanced components as $g_{\rm a}^{\rm K}(k,\omega,\theta) = \left( g_{\rm a}^{\rm ret}(k,\omega,\theta) - g_{\rm a}^{\rm adv}(k,\omega,\theta) \right) ( 1 + 2n(\omega,T_{\rm a}))$, where $T_{\rm a}$ is the temperature of material $\mathrm{a}$ and $n(\omega,T_{\rm a})$ is the Bose-Einstein distribution function.
The retarded and advanced components of the Green's function are given by the recursive Green's function method for a semi-infinite system.

\textit{Results and discussion.}---
We now evaluate the spin current driven by the temperature difference $\Delta T = T_{\rm L} - T_{\rm R}$ at the mean temperature $\overline{T} = (T_{\rm L} + T_{\rm R})/2$ using Eq.~\eqref{eq:Keldysh-Green}.
To remove the trivial scalings, we express the spin current in a dimensionless form as
\begin{align}
    g
    =
    \frac{\pi}{k_{\rm c} \Delta T}
    \left(
        \frac{2J^2}{(J^{\rm ex}_{\rm AA})^{2} + (J^{\rm ex}_{\rm BB})^{2}}
    \right)
    j_{S_{z}}(\theta).
    \label{eq:dimensionless-spin-current}
\end{align}
To leading order in the interface coupling, $j_{S_{z}}(\theta)$ is quadratic in $V$ and involves the surface spectral functions of the two magnets whose energy scale is the magnon bandwidth $JS$, so that $[(J^{\rm ex}_{\rm AA})^{2} + (J^{\rm ex}_{\rm BB})^{2}]/2J^{2}$ is the natural measure of the coupling.
The response is linear in $\Delta T$, and since the momentum $k$ parallel to the interface is conserved, each $k$ in $|k| \leq k_{\rm c}$ is an independent tunneling channel, $k_{\rm c}/\pi$ of them per unit interface length.
All results below are computed at $\eta = JS/32$ and $k_{\rm c} = 1/a$ unless stated otherwise; the symmetry relations discussed below do not depend on them, whereas the absolute scale of $g$ does.

\begin{figure}
    \centering 
    \includegraphics[scale=0.8]{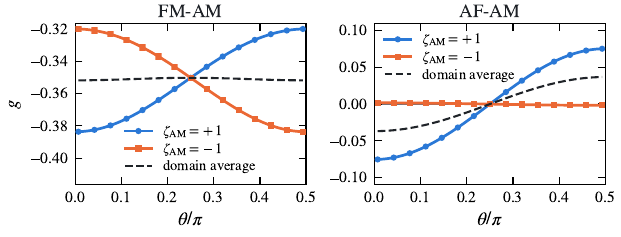}
    \caption{The dimensionless spin current $g$ as a function of the interface angle $\theta$.
    The left and right panels show the results for the FM/AM and AFM/AM junctions, respectively.
    The blue and red curves with circles and squares show the results for magnetic domains of altermagnet $\zeta_{\rm AM} = 1$ and $\zeta_{\rm AM} = -1$, respectively.
    The black dashed lines are the domain-averaged values.
    $K = 0.05J$, $\delta J = 0.1J$, $S = 1$, $J^{\rm ex}_{\rm AA} = J^{\rm ex}_{\rm BB} = -0.1J$, hence $\mathcal{J}_{\rm AA} = \mathcal{J}_{\rm BB} = +0.1J$, $k_{\rm c} = 1/a$, $\overline{T} = JS$, $\Delta T = 0.4JS$, $\eta = JS/32$.
    $J_0 = 0.1J$ for the ferromagnetic case, and $J_0 = -0.1J$ for the antiferromagnetic and altermagnetic cases.}
    \label{fig:g_angle}
\end{figure}

Figure~\ref{fig:g_angle} shows $g(\theta)$ for the symmetric coupling $J^{\rm ex}_{\rm AA} = J^{\rm ex}_{\rm BB}$, for the two domains $\zeta_{\rm AM} = \pm 1$ of the altermagnet and their average $\bar{g}$.
The domain average $\bar{g}(\theta) = [g(\theta; +1) + g(\theta; -1)]/2$ is the equal-weight average over the two domains of the altermagnet at a fixed N\'eel vector of the antiferromagnet (magnetization of the ferromagnet).
The two domains are related by time reversal and, equivalently, by the fourfold rotation of the altermagnetic anisotropy, $\delta J \to -\delta J$, since the sublattice exchange maps $(\zeta_{\rm AM}, \delta J)$ onto $(-\zeta_{\rm AM}, -\delta J)$; they exhaust the domains of a single-crystal $d$-wave altermagnet, and domain walls are not considered.
Here, $g > 0$ denotes $S_{z}$ flowing into the altermagnet; the FM/AM current is negative because the hotter ferromagnet injects magnons carrying $S_{z} = -1$.
All curves take the form $g(\theta) = c_{0} + c_{2}\cos 2\theta$, but the two junctions differ qualitatively.
In the FM/AM junction, a $\cos 2\theta$ modulation of about $9\%$ rides on a uniform background, $c_{0}$ is even and $c_{2}$ is odd under reversal of $\zeta_{\rm AM}$, and thus the two domains are mirror images of each other about $\theta = \pi/4$ and the anisotropy cancels on the domain average.
In the AFM/AM junction, $c_{0} = 0$ and the current for $\zeta_{\rm AM} = +1$ is a pure $\cos 2\theta$ with a node at $\theta = \pi/4$; the current for $\zeta_{\rm AM} = -1$ is a pure $\cos 2\theta$ as well but suppressed to $c_{2}^{-}/c_{2}^{+} \simeq -0.02$, where $c_{n}^{\pm}$ denotes $c_{n}$ for $\zeta_{\rm AM} = \pm 1$, and thus the $d$-wave anisotropy survives the domain average.

These features follow from the way $\theta$ enters the junction.
Since the interface couples only the same sublattices, $\theta$ enters solely through the altermagnetic term of Eq.~\eqref{eq:spin-wave-Hamiltonian-AF}, whose long-wavelength form is
\begin{align}
    \delta\gamma_{\rm A} = -\delta\gamma_{\rm B} = \delta J S a^2 \left\{ \left( k_{\parallel}^2 - k_{\perp}^2 \right) \cos2\theta - 2 k_{\parallel} k_{\perp} \sin2\theta \right\},
    \label{eq:anisotropic-term}
\end{align}
where $k_{\parallel}$ and $k_{\perp}$ are the momenta along $\vb{k}_{\parallel}$ and $\vb*{n}$, respectively.
Equation~\eqref{eq:anisotropic-term} is invariant under $(\theta, k_{\parallel}) \to (-\theta, -k_{\parallel})$ and $\pi$-periodic in $\theta$, so that $j_{S_{z}}(\theta) = \sum_{n \geq 0} c_{2n} \cos 2n\theta$; the higher harmonics $n \geq 2$, at most $3\%$ of $c_{2}$ \cite{SM}, are neglected.
Moreover, Eq.~\eqref{eq:anisotropic-term} changes its sign under $(\theta, k_{\parallel}) \to (\pi/2 - \theta, -k_{\parallel})$, so that the reflection of the interface angle about $\pi/4$ is equivalent to the reversal of the altermagnetic order parameter,
\begin{align}
    j_{S_{z}}(\theta; \delta J) = j_{S_{z}}(\pi/2 - \theta; -\delta J).
    \label{eq:reflection}
\end{align}
The two junctions differ in whether this reversal can be traded for a symmetry operation.
Let $(\pm, \pm)$ denote the signs of $S_{z}$ on the (A, B) sublattices; the antiferromagnet and the altermagnet with $\zeta_{\rm AM} = +1$ are $(+,-)$, and the ferromagnet is $(+,+)$.
For the FM/AM junction, the sublattice exchange $\mathrm{A} \leftrightarrow \mathrm{B}$ leaves the ferromagnet invariant and reverses both $\delta J$ and $\zeta_{\rm AM}$ in the altermagnet, giving with Eq.~\eqref{eq:reflection} $j_{S_{z}}(\theta; \zeta_{\rm AM}) = j_{S_{z}}(\pi/2 - \theta; -\zeta_{\rm AM})$, i.e. $c_{0}$ is even and $c_{2}$ is odd under reversal of $\zeta_{\rm AM}$.
For the AFM/AM junction with $J^{\rm ex}_{\rm AA} = J^{\rm ex}_{\rm BB}$, the spin rotation $S_{z} \to -S_{z}$ followed by $\mathrm{A} \leftrightarrow \mathrm{B}$ maps $(+,-)$ of both magnets onto themselves while reversing the sign of the spin current and of $\delta J$, giving with Eq.~\eqref{eq:reflection} $j_{S_{z}}(\theta) = -j_{S_{z}}(\pi/2 - \theta)$, i.e. $c_{0} = 0$ and the node at $\theta = \pi/4$ within one domain.
Since the global spin rotation $S_{z} \to -S_{z}$ alone reverses the current and both N\'eel vectors, $j_{S_{z}}$ is odd under the simultaneous reversal of the two N\'eel vectors: the surviving anisotropy is a property of the relative N\'eel configuration, the domain average over the altermagnet must be taken at a fixed N\'eel vector of the antiferromagnet, and a multidomain antiferromagnet with equal weights would cancel it.
The selection rule concerns the spin connected by the interface bond, i.e. for the cross-sublattice channel, the transmitting domain is $\zeta_{\rm AM} = -1$.
When both couplings coexist with $\mathcal{J}_{\rm AA} = \mathcal{J}_{\rm BB}$ and $\mathcal{J}_{\rm AB} = \mathcal{J}_{\rm BA}$, $c_{0}$ still vanishes.
The anisotropic component $c_{2}$ survives the domain average unless the two couplings coexist, of which the detail is given in the SM~\cite{SM}.

\begin{figure}
    \centering 
    \includegraphics[scale=0.8]{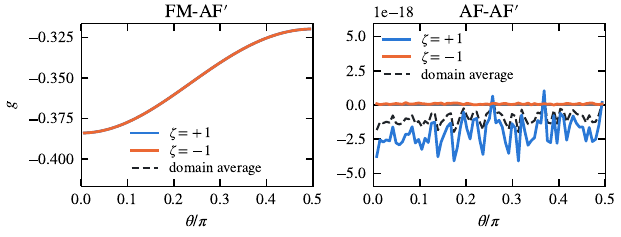}
    \caption{The interface angle dependence of the dimensionless spin current $g$ for the anisotropic antiferromagnetic system.
    The left panel shows the result for a junction with a ferromagnet, and the right panel shows the result with an antiferromagnet.
    The black dashed lines are the domain-averaged values.
    $K = 0.05J$, $\delta J = 0.1J$, $S = 1$, $J^{\rm ex}_{\rm AA} = J^{\rm ex}_{\rm BB} = -0.1J$, hence $\mathcal{J}_{\rm AA} = \mathcal{J}_{\rm BB} = +0.1J$, $k_{\rm c} = 1/a$, $\overline{T} = JS$, $\Delta T = 0.4JS$, $\eta = JS/32$.
    $J_0 = 0.1J$ for the ferromagnetic case, and $J_0 = -0.1J$ for the antiferromagnetic cases.
    }
    \label{fig:g_angle_afmp}
\end{figure}

To verify the characteristics of the altermagnet, we compare with the anisotropic antiferromagnetic (denoted as AFM') system, which is described by the sublattice-even anisotropy introduced by the intra-sublattice exchange $J \mp \delta J$ along the $x$ and $y$ directions on both sublattices.
The results for junction systems FM/AFM' and AFM/AFM' are shown in Fig.~\ref{fig:g_angle_afmp}.
In the FM/AFM' junction, the magnetic domain dependence disappears.
This is because the domain reversal of AFM' can be absorbed by the relabeling of the sublattices, and thus the domain reversal pairs are equivalent when $\mathcal{J}_{\rm AA} = \mathcal{J}_{\rm BB}$.
In the AFM/AFM' junction, the isotropic component $c_{0}$ vanishes by the same symmetry as in the case of the AFM/AM junction.
Furthermore, the anisotropic component $c_{2}$ also vanishes by the combination of spin reversal $S_{z} \to -S_{z}$ and sublattice relabeling, which leads to $j_{S_{z}}(\theta; \delta J) = -j_{S_{z}}(\theta; \delta J)$.
As a result, the spin current in the AFM/AFM' junction is zero up to the numerical error scale.
Therefore, we can conclude that the nodal structure of the spin current in the AFM/AM junction is unique to the altermagnetic order.

\begin{figure}
    \centering 
    \includegraphics[width=0.72\columnwidth]{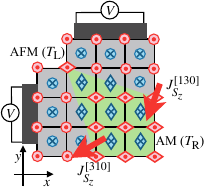}
    \caption{Proposed lateral AFM/AM junctions with the compensated interfaces normal to $[310]$ and $[130]$, related by the reflection $\theta \to \pi/2 - \theta$, under the temperature difference $\Delta T = T_{\rm L} - T_{\rm R}$ between the AFM and the AM.
    The interface spin current is read by the inverse spin Hall voltage of a heavy-metal electrode on the AFM; the two interfaces give equal and opposite currents.}
    \label{fig:exp}
\end{figure}

\textit{Proposal for an experiment}---
To end this letter, we propose two measurements on a lateral AFM/AM junction (Fig.~\ref{fig:exp}).
The first detects the $d$-wave anisotropy: for the compensated interfaces normal to $[310]$ and $[130]$, at $\tan\theta = 1/3$ and $\tan\theta = 3$ and related by the reflection $\theta \to \pi/2 - \theta$, the currents are equal and opposite, $j^{[310]}_{S_{z}} = -j^{[130]}_{S_{z}} = 0.8\,c_{2}$, so that their difference is the signal and their sum is a null test for any uniform component; the same pair in an FM/AM junction gives a finite sum $2c_{0}$ and a difference that vanishes on the domain average.
The second reads the N\'eel vector of the altermagnet: at a fixed interface the current is suppressed when the altermagnetic domain is reversed relative to the N\'eel vector of the antiferromagnet, for $|J_{0}| = 0.1J$ ($J$), and a multidomain altermagnet gives the domain fraction.
For a lattice continuous across the interface, the nearest-neighbor bonds crossing a compensated plane connect opposite sublattices, so that the transmitting configuration is the one in which the bonded spins are parallel; which configuration is the ground state is decided by the sign of the interface exchange, whereas the sign reversal between $[310]$ and $[130]$ and the valve readout at a fixed interface do not depend on the registry.
The interface spin current can be detected as the inverse spin Hall voltage of a heavy-metal electrode on the antiferromagnet, as in nonlocal magnon transport in antiferromagnetic insulators \cite{Magnon-tunneling_ref-in-paper_nonlocal-AFM_2018_Lebrun}; if the injected $S_{z}$ acts as a spin-transfer torque of opposite sign on the two circular modes of the antiferromagnet \cite{Magnon-tunneling_ref-in-paper_AFM-STT_2014_Cheng}, the sign of the current also appears as the shifts $\delta\Gamma_{\pm}$ of the two circular AFMR linewidths, with $\delta\Gamma_{+} = -\delta\Gamma_{-}$ \cite{Magnon-tunneling_ref-in-paper_AFMR_2020_Vaidya}.

\textit{Conclusion.}---
In summary, we have identified distinct signatures of altermagnetic order in thermally driven magnon tunneling across FM/AM and AFM/AM junctions with sublattice-symmetric interfacial exchange interaction.
In the FM/AM junction, the AM domain average retains a finite interface angle-independent spin current but eliminates its anisotropic contribution.
In the AFM/AM junction, symmetry forbids the uniform contribution, while the relative spin alignment of the coupled interfacial sublattices gives rise to the magnon spin valve effect. 
By comparing with the anisotropic antiferromagnet, we find that the anisotropy cancellation in the FM/AM junction and the finite anisotropic spin current in the AFM/AM junction are unique to the altermagnet due to the sublattice-odd anisotropic exchange interaction.

Our description with localized spins on the lattice sites is most directly applicable to magnetic insulator junctions.
Extensions to metallic systems require accounting for itinerant-electron contributions and their exchange coupling to localized spin moments, while the effects of spin-orbit coupling remain to be assessed. 
These results provide a basis for probing altermagnetic anisotropy in domain-controlled samples and suggest a Joule-heating-free electrical readout of the N\'eel alignment of the AM relative to an AFM reference through magnon spin transport.

\textit{Data availability.}---
The data used in this letter will be deposited in the Nagoya University Research Information Repository~\cite{Magnon-tunneling_Nagoya-Repository} upon publication.

\textit{Acknowledgments.}---
The author is grateful to Ai Yamakage and Yuta Sekino for discussions and advice during the preparation of this letter.
This work was supported by JST SPRING (Grant No. JPMJSP2125).

\bibliography{ref}

\clearpage
\onecolumngrid
\begin{center}
    \textbf{Supplemental Material}
\end{center}

\section{Junction system and magnon tunneling}
\subsection{Model Hamiltonian and spin-wave approximation}
We start with the Heisenberg Hamiltonian for the left and right materials, which is given by
\begin{align}
    H_{\rm a}
    =&
    -\sum_{n \in \mathbb{Z}} \sum_{l,l' \in \{A,B\}} \sum_{\vec{\delta}}
    J^{\rm a}_{n,\vec{\delta},l,l'}
    \vb{S}^{\rm a}_{\vec{r}_{n,l}} \cdot \vb{S}^{\rm a}_{\vec{r}_{n,l'}+\vec{\delta}}
    -
    K \sum_{n \in \mathbb{Z}} \sum_{l \in \{A,B\}}
    \left( S^{\rm a}_{z,\vec{r}_{n,l}} \right)^2,
\end{align}
where $\vb{S}^{\rm a}_{\vec{r}_{n,l}}$ is the spin operator at site $\vec{r}_{n,l}$, $J^{\rm a}_{n,\vec{\delta},l,l'}$ is the exchange interaction strength between the spins at sites $\vec{r}_{n,l}$ and $\vec{r}_{n,l'}+\vec{\delta}$, and $K$ is the easy-axis anisotropy constant.
In what follows, we introduce the Holstein--Primakoff transformation and spin-wave approximation to describe the low-energy excitations from the ferromagnetic, antiferromagnetic, and altermagnetic ground states.
For all three cases, we put the ferromagnetic interaction between the same sublattices as $J^{\rm a}_{n,\vec{\delta},l,l} > 0$.
Similarly, for the ferromagnetic case, we give the condition $J^{\rm a}_{n,\vec{\delta},l,l'} =: J_{0} > 0$ for the interaction between different sublattices, while for the antiferromagnetic and altermagnetic cases, we give the condition $J_{0} < 0$.
The difference between the antiferromagnetic and altermagnetic cases is the anisotropy of the exchange interaction between the same sublattices.
Here we denote this anisotropy as $\delta J$; it has opposite signs on the two sublattices, $J^{\rm a}_{n,\vec{\delta},l,l} = J - \sigma_{l}\delta J$ for $\vec{\delta} = \pm a \hat{x}$ and $J^{\rm a}_{n,\vec{\delta},l,l} = J + \sigma_{l}\delta J$ for $\vec{\delta} = \pm a \hat{y}$ with $\sigma_{A} = -\sigma_{B} = 1$, so that the two sublattices are related by the $C_{4}$ rotation.

\subsection{Ferromagnetic case}
First, we consider the ferromagnetic case.
The Holstein--Primakoff transformation is given by
\begin{align}
    S^{\rm a}_{+,\vec{r}_{n,l}} = \sqrt{2S} \phi_{\vec{r}_{n,l}}, \quad
    S^{\rm a}_{-,\vec{r}_{n,l}} = \phi_{\vec{r}_{n,l}}^\dag \sqrt{2S}, \quad
    S^{\rm a}_{z,\vec{r}_{n,l}} = S - \phi_{\vec{r}_{n,l}}^\dag \phi_{\vec{r}_{n,l}},
\end{align}
where $\phi_{\vec{r}_{n,l}}$ and $\phi_{\vec{r}_{n,l}}^\dag$ are the bosonic annihilation and creation operators at site $\vec{r}_{n,l}$, respectively, satisfying the commutation relation $[\phi_{\vec{r}_{n,l}}, \phi_{\vec{r}_{n',l'}}^\dag] = \delta_{n,n'}\delta_{l,l'}$.
The constant $S$ is the spin quantum number.
To describe the low-energy excitations, we keep only the quadratic terms in the bosonic operators, which leads to the spin-wave Hamiltonian.
With the Fourier transformation $\phi_{\vec{r}_{n,l}} = \frac{1}{\sqrt{N}} \sum_{\vec{k}} e^{i\vec{k}\cdot\vec{r}_{n,l}} \phi_{\vec{k},l}$, where $N$ is the number of lattice sites, we can rewrite the spin-wave Hamiltonian in momentum space as
\begin{align}
    H_{\rm a}^{\rm F}
    = \sum_{\vec{k}} \mqty( \phi_{\vec{k}, A}^{\dag} & \phi_{\vec{k}, B}^{\dag} )
    \mqty( 
        \gamma_{\rm F}(\vec{k}) & \gamma_{0}(\vec{k}) \\
        \gamma_{0}(\vec{k}) & \gamma_{\rm F}(\vec{k})
    )
    \mqty( \phi_{\vec{k}, A} \\ \phi_{\vec{k}, B} ),
\end{align}
where $\gamma_{\rm F}(\vec{k}) = (4J_{0} + 2K)S + 2JS \sum_{\mu} (1 - \cos(ak_{\mu}))$ and $\gamma_{0}(\vec{k}) = -4J_{0}S \cos(ak_x/2)\cos(ak_y/2)$.
To describe the direction of the interface as a continuous variable, we use the layer coefficients of the numerical Hamiltonian,
\begin{align}
    &\gamma_{\rm F}(\vec{k}) \simeq \gamma_{\rm F}'(k_{\parallel}) + JS a^2 k_{\perp}^2, \quad
    \gamma_{0}(\vec{k}) \simeq \gamma_{0}'(k_{\parallel}) + \frac{J_{0} S a^2}{2}k_{\perp}^2,
    \notag\\
    &\gamma_{\rm F}'(k_{\parallel}) = (4J_{0} + 2K)S + JS a^2 k_{\parallel}^2, \quad
    \gamma_{0}'(k_{\parallel}) = -4J_{0}S + \frac{J_{0} Sa^2}{2 }k_{\parallel}^2,
\end{align}
where $k_{\perp} = k_x \cos\theta + k_y \sin\theta$ and $k_{\parallel} = k_x \sin\theta - k_y \cos\theta$ are the momentum components along the interface normal $\vb*{n} = (\cos\theta, \sin\theta)$ and along the interface direction $(\sin\theta, -\cos\theta)$, respectively, and $\theta$ is the angle between $\vb*{n}$ and the $x$-axis.
The long-wavelength coefficient along the interface used in the numerics is $JSa^{2}k_{\parallel}^{2}$, which is the parallel piece of the isotropic expansion of $2JS\sum_{\mu}(1-\cos(ak_{\mu}))$.
According to the central difference approximation, the discretization of the perpendicular momentum is given by
\begin{align}
    k_{\perp}^2 \phi_{\vec{k}, l} \simeq \frac{2\phi_{k_{\parallel}, j, l} - \phi_{k_{\parallel}, j+1, l} - \phi_{k_{\parallel}, j-1, l}}{a_{\perp}^2},
\end{align}
where $a_{\perp}$ is the lattice constant in the perpendicular direction.
In the numerical calculations, we use $a_{\perp} = 1.0$.
Thus, the spin-wave Hamiltonian can be rewritten as
\begin{align}
    H^{\rm F}_{\rm L} 
    &= \sum_{k_{\parallel}}\sum_{j < 0}
    \left[
        \phi_{k_{\parallel},j}^{\dag} \hat{h}^{\rm F}_{k_\parallel} \phi_{k_{\parallel}, j}
        +
        \phi_{k_{\parallel},j}^{\dag} \hat{t}^{\rm F}_{k_{\parallel}} \phi_{k_{\parallel}, j+1}
        +
        \phi_{k_{\parallel},j+1}^{\dag} \hat{t}^{\rm F \dag}_{k_{\parallel}} \phi_{k_{\parallel}, j}
    \right]
    \notag\\ 
    \hat{h}^{\rm F}_{k_{\parallel}} &= \mqty(
        \gamma_{\rm F}'(k_{\parallel}) + \frac{2JS a^2}{a_{\perp}^2} & \gamma_{0}'(k_{\parallel}) + \frac{J_{0} S a^2}{ a_{\perp}^2} \\
        \gamma_{0}'(k_{\parallel}) + \frac{J_{0} S a^2}{ a_{\perp}^2} & \gamma_{\rm F}'(k_{\parallel}) + \frac{2JS a^2}{a_{\perp}^2}
    ),
    \quad 
    \hat{t}^{\rm F}_{k_{\parallel}} = - \mqty(
        \frac{JS a^2}{a_{\perp}^2} & \frac{J_{0} S a^2}{2 a_{\perp}^2} \\
        \frac{J_{0} S a^2}{2 a_{\perp}^2} & \frac{JS a^2}{a_{\perp}^2}
    ),
    \quad 
    \phi_{k_{\parallel}, j} = \mqty( \phi_{k_{\parallel}, j, A} \\ \phi_{k_{\parallel}, j, B} ).
\end{align}

\subsection{Antiferromagnetic and altermagnetic cases}
Next, we consider the antiferromagnetic and altermagnetic cases.
The Holstein--Primakoff transformation is written with the canonical magnon $a^{\rm a}_{\vec{r}_{n,l},l}$ defined about the classical ground state of the magnetic domain $\zeta_{\rm a} = \pm 1$, in which the moment on the sublattice $l$ points along $\zeta_{\rm a}\sigma_{l}$ with $\sigma_{A} = -\sigma_{B} = 1$:
\begin{align}
    S^{\rm a}_{+,\vec{r}_{n,l}} = \sqrt{2S} \psi^{\rm a}_{\vec{r}_{n,l},l}, \quad
    S^{\rm a}_{-,\vec{r}_{n,l}} = \psi^{\rm a \dag}_{\vec{r}_{n,l},l} \sqrt{2S}, \quad
    S^{\rm a}_{z,\vec{r}_{n,l}} = \zeta_{\rm a}\sigma_{l} \lb S - n^{\rm a}_{\vec{r}_{n,l},l} \rb,
    \quad
    \psi^{\rm a}_{\vec{r}_{n,l},l} =
    \begin{cases}
        a^{\rm a}_{\vec{r}_{n,l},l} & (\zeta_{\rm a}\sigma_{l} = +1) \\
        a^{\rm a \dag}_{\vec{r}_{n,l},l} & (\zeta_{\rm a}\sigma_{l} = -1)
    \end{cases},
\end{align}
where $n^{\rm a}_{\vec{r}_{n,l},l} = a^{\rm a \dag}_{\vec{r}_{n,l},l} a^{\rm a}_{\vec{r}_{n,l},l}$ is the number operator of the canonical magnon, and the field $\psi^{\rm a}_{\vec{r}_{n,l},l} = S^{\rm a}_{+,\vec{r}_{n,l}}/\sqrt{2S}$ is the spin-raising operator, which is the annihilation operator of the magnon on a sublattice whose moment points up and the creation operator on a sublattice whose moment points down.
The commutator of the field follows as $[\psi^{\rm a}_{\vec{r}_{n,l},l}, \psi^{\rm b \dag}_{\vec{r}_{n',l'},l'}] = \zeta_{\rm a} \sigma_{l} \delta_{\rm a, \rm b} \delta_{n,n'} \delta_{l, l'}$.
The bosonic BdG spinor used below, $\Psi = (\psi_{A}, \psi_{B})^{\rm T}$, therefore has the metric $[\Psi, \Psi^{\dag}] = \zeta_{\rm a}\Sigma_{z}$ with $\Sigma_{z} = \mathrm{diag}(\sigma_{A}, \sigma_{B}) = \mathrm{diag}(1, -1)$, which is derived from the local Holstein--Primakoff rule rather than postulated.
We use the spin-raising field rather than the canonical magnon as the component of the spinor because the sublattice order, the Hamiltonian matrix, and the sublattice-diagonal interface coupling are then the same in both domains; the domain enters only through the metric $\zeta_{\rm a}\Sigma_{z}$.
This is also the basis of the numerical implementation, in which the left electrode is kept in its reference domain and only the altermagnet is reversed.
In this paper, we introduce the magnetic domain dependence only for the altermagnetic material, $\zeta_{\rm R} = \zeta_{\rm AM}$, and we set $\zeta_{\rm L} = 1$ for the ferromagnetic and antiferromagnetic materials.
According to the Fourier transformation $\psi^{\rm a}_{\vec{r}_{n,l},l} = \frac{1}{\sqrt{N}} \sum_{\vec{k}} e^{i\vec{k}\cdot\vec{r}_{n,l}} \psi^{\rm a}_{\vec{k},l}$ of the spin-raising field, for which $\psi^{\rm a}_{\vec{k},B} = a^{\rm a \dag}_{-\vec{k},B}$ in the reference domain, we can rewrite the spin-wave Hamiltonian in momentum space as
\begin{align}
    H_{\rm a}^{\rm AF}
    = \sum_{\vec{k}} \mqty( \psi_{\vec{k}, A}^{\rm a \dag} & \psi^{\rm a \dag}_{\vec{k}, B} )
    \mqty(
        \gamma_{\rm A}(\vec{k}) & \gamma_{0}(\vec{k}) \\
        \gamma_{0}(\vec{k}) & \gamma_{\rm B}(\vec{k})
    )
    \mqty( \psi^{\rm a}_{\vec{k}, A} \\ \psi^{\rm a}_{\vec{k}, B} ),
\end{align}
where $\gamma_{\rm A}(\vec{k}) = (-4J_{0} + 2K)S + 2JS \sum_{\mu} (1 - \cos(ak_{\mu})) + 2\delta J S (\cos(ak_x) - \cos(ak_y))$ and $\gamma_{\rm B}(\vec{k}) = (-4J_{0} + 2K)S + 2JS \sum_{\mu} (1 - \cos(ak_{\mu})) - 2\delta J S (\cos(ak_x) - \cos(ak_y))$.
By the bosonic Bogoliubov transformation, the spin-wave spectrum is obtained as 
\begin{align}
    \omega_{\pm, \vec{k}}
    = \sqrt{ (\gamma_{\rm AF}(\vec{k}))^2 -  (\gamma_{0}(\vec{k}))^2 } \mp 2S\delta J  \left( \cos(ak_x) - \cos(ak_y) \right),
    \quad 
    \gamma_{\rm AF}(\vec{k}) = (-4J_{0} + 2K)S + 2JS \sum_{\mu} (1 - \cos(ak_{\mu})) 
\end{align}
Here, the long-wavelength limit of the term of $\delta J$ is given as 
\begin{align}
    2\delta J S \left[ \cos(ak_x) - \cos(ak_y) \right]
    \simeq 
    \delta J S a^{2}
    \left\{
        \left( k_{\parallel}^2 - k_{\perp}^2 \right) \cos 2\theta - 2k_{\parallel}k_{\perp}\sin2\theta
    \right\}.
\end{align}
The discretization of the linear momentum is given by
\begin{align}
    k_{\perp} \psi^{\rm a}_{\vec{k}, l} \simeq -i\,\frac{\psi^{\rm a}_{k_{\parallel}, j+1, l} - \psi^{\rm a}_{k_{\parallel}, j-1, l}}{2 a_{\perp}}.
\end{align}
Thus, the spin-wave Hamiltonian can be rewritten as
\begin{align}
    H^{\rm AF/AM}_{\rm a} 
    &= \sum_{k_{\parallel}}{\sum_{j}}^{\rm a}
    \left[
        \psi_{k_{\parallel},j}^{\rm a\dag} \hat{h}^{\rm AF/AM}_{k_\parallel} \psi^{\rm a}_{k_{\parallel}, j}
        +
        \psi_{k_{\parallel},j}^{\rm a\dag} \hat{t}^{\rm AF/AM}_{k_{\parallel}} \psi^{\rm a}_{k_{\parallel}, j+1}
        +
        \psi_{k_{\parallel},j+1}^{\rm a\dag} \hat{t}^{\rm AF/AM \dag}_{k_{\parallel}} \psi^{\rm a}_{k_{\parallel}, j}
    \right]
    \notag\\ 
    \hat{h}^{\rm AF}_{k_{\parallel}} =& 
    \hat{h}^{\rm AF}_{0, k_{\parallel}}
    +
    S\delta J a^{2} \lb k_{\parallel}^2  - \frac{2}{a_{\perp}^2} \rb \cos 2\theta
    \mqty(
        1 & 0 \\
        0 & -1 
    ),
    \quad 
    \hat{t}^{\rm AF}_{k_{\parallel}} = \hat{t}^{\rm F}_{k_{\parallel}} + 
    \left(
        \frac{S\delta J a^2 }{a_\perp^{2}} \cos 2\theta
        +
        i \frac{S\delta J a^2 k_{\parallel}}{a_\perp} \sin 2\theta
    \right) \mqty( 1 & 0 \\ 0 & -1 ),
    \notag\\
    \hat{h}^{\rm AF}_{0, k_{\parallel}} &= 
    \mqty(
        \left. \gamma_{\rm F}'(k_{\parallel}) \right|_{J_{0} \to -J_{0}} + \frac{2JS a^2}{a_{\perp}^2} & \gamma_{0}'(k_{\parallel}) + \frac{J_{0} S a^2}{ a_{\perp}^2} \\
        \gamma_{0}'(k_{\parallel}) + \frac{J_{0} S a^2}{ a_{\perp}^2} & \left. \gamma_{\rm F}'(k_{\parallel}) \right|_{J_{0} \to -J_{0}} + \frac{2JS a^2}{a_{\perp}^2}
    ),
    \notag\\
    \psi_{k_{\parallel}, j} =& \mqty( \psi_{k_{\parallel}, j, A} \\ \psi_{-k_{\parallel}, j, B}^{\dag} ),
\end{align}
where the summation ${\sum_{j}}^{\rm a}$ is taken over $j < 0$ for the left material and $j \geq 0$ for the right material.
The letter's $\Psi_{\vec{k}}^{\rm a}$ is the bulk spinor, while the layer spinor $\psi_{k_{\parallel},j}$ defined here is its BdG representation.

\subsection{Spin current operator and Keldysh formalism}
This section presents the derivation of the spin current operator and its expectation value using the Keldysh formalism.
The spin current operator is defined as the time derivative of the spin operator on the right-hand side of the interface, which is given by
\begin{align}
    \hat{I}_{S_z}^{\rm R} 
    = \dv{\vb*{S}_\mathrm{R}}{t} 
    = \frac{1}{i} [\vb*{S}_\mathrm{R}, H_\mathrm{I}]
    = \sum_{\langle i,j \rangle}\frac{\mathcal{J}_{i,j}}{2i}
        \left(
S_{+,i}^{\rm L} S_{-,j}^{\rm R} - S_{-,i}^{\rm L} S_{+,j}^{\rm R}
        \right),
\end{align}
where $\vb*{S}_\mathrm{R} = \sum_{i \in \mathrm{interface}} \vb*{S}_i^{\rm R}$ is the total spin operator on the right side of the interface, and $S_{\pm,i}^{\rm a} = S_{x,i}^{\rm a} \pm i S_{y,i}^{\rm a}$ are the spin raising and lowering operators, which are simplified expressions of the original spin operators in the above section.
By using the same definition, we can find that $\hat{I}_{S_z}^{\rm R} = - \hat{I}_{S_z}^{\rm L}$, where $\hat{I}_{S_z}^{\rm L}$ is the spin current operator on the left side of the interface.
In the Keldysh formalism, the expectation value of the spin current operator is given by 
\begin{align}
    \langle \hat{I}_{S_z}^{\rm R} \rangle
    = \sum_{\langle i,j \rangle} \mathcal{J}_{i,j}\,
        \mathrm{Im}
        \langle  S_{+,i}^{\rm L} S_{-,j}^{\rm R} \rangle.
\end{align}
The value that enters the numerical current is not this raw expectation. With the Holstein--Primakoff hopping $J^{\rm ex}_{\mu\nu}=-\sqrt{S_{\rm L}S_{\rm R}}\,\mathcal{J}_{\mu\nu}$ and the boundary-spinor Green's function $G_{\rm LR}=-i\langle T_{\mathcal C}\chi_{\rm L}\chi_{\rm R}^{\dagger}\rangle$,
\begin{align}
    \langle \hat{I}_{S_z}^{\rm R} \rangle
    =
    -\mathrm{Re}\,\mathrm{Tr}_{\rm R}\!\left[V_{\rm RL}G_{\rm LR}^{\rm K}\right],
\end{align}
where $V_{\rm RL}$ is the Hermitian conjugate of the hopping matrix $V$ built from $J^{\rm ex}_{\mu\nu}$. A positive $\mathcal{J}_{\mu\nu}$ is ferromagnetic. The equal-time contraction above is the one integrated in the main text.
The Fourier transformed retarded, advanced and Keldysh components denoted by $G^{\rm ret/adv/K}(k,\omega)$ can be written in matrix form as 
\begin{align}
    G(k,\omega) = \mqty(
        G^{\rm ret}(k,\omega) & G^{\rm K}(k,\omega) \\
        0 & G^{\rm adv}(k,\omega)
    ).
\end{align}
With the interaction matrix $V$, the Fourier space counterpart of $J^{\rm ex}_{i,j}$ for the magnon exchange term, the Dyson equation for the correlation function is given by
\begin{align}
    G(k, \omega) = g_{\rm L}(k, \omega) V g_{\rm R}(k, \omega) \left( I + V G(k, \omega) \right),
\end{align}
where $g_{\rm L/R}(k, \omega)$ is the Green's function of the left/right material; explicit forms of this Green's function are given in the following section.
Thus, the correlation function can be expressed as
\begin{align}
    G(k, \omega) = \left[ \left( g_{\rm L} V g_{\rm R} \right)^{-1} - V \right]^{-1}.
\end{align}
In this equation, the first term is obtained as 
\begin{align}
   \left( g_{\rm L} V g_{\rm R} \right)^{-1}
   &=
   \mqty(
        g_{\rm L}^{\rm ret}(k, \omega) V g_{\rm R}^{\rm ret}(k, \omega) & F_{\rm L, R}(k, \omega) \\
        0 & g_{\rm L}^{\rm adv}(k, \omega) V g_{\rm R}^{\rm adv}(k, \omega) 
   )^{-1}
   \notag\\
   &=
    \mqty(
          \left( G_{\rm L,R}^{\rm r,r}(k, \omega) \right)^{-1} & -\left( G_{\rm L,R}^{\rm r,r}(k, \omega) \right)^{-1} F_{\rm L, R}(k, \omega) \left( G_{\rm L,R}^{\rm a,a}(k, \omega) \right)^{-1} \\
          0 & \left( G_{\rm L,R}^{\rm a,a}(k, \omega) \right)^{-1}
    ),
\end{align}
where $G_{\rm L,R}^{\rm r,r}(k, \omega) = g_{\rm L}^{\rm ret}(k, \omega) V g_{\rm R}^{\rm ret}(k, \omega)$, $G_{\rm L,R}^{\rm a,a}(k, \omega) = g_{\rm L}^{\rm adv}(k, \omega) V g_{\rm R}^{\rm adv}(k, \omega)$, and $F_{\rm L, R}(k, \omega) = g_{\rm L}^{\rm ret}(k, \omega) V g_{\rm R}^{\rm K}(k, \omega) + g_{\rm L}^{\rm K}(k, \omega) V g_{\rm R}^{\rm adv}(k, \omega)$.
The Keldysh component satisfies the fluctuation-dissipation theorem in the form 
\begin{align}
    g_{\rm L/R}^{\rm K}(k, \omega) = \left( g_{\rm L/R}^{\rm ret}(k, \omega) - g_{\rm L/R}^{\rm adv}(k, \omega) \right) \left( 1 + 2 n(\omega; T_{\rm L/R}) \right),
\end{align}
where $n(\omega; T) = 1/(e^{\omega/T}-1)$ is the Bose distribution function, and $T_{\rm L/R}$ is the temperature of the left/right material.
Therefore, the Keldysh component of the correlation function is given by
\begin{align}
    G^{\rm K}(k, \omega)
    &= \left( I - G_{\rm L,R}^{\rm r,r} V \right)^{-1} F_{\rm L, R}(k, \omega) \left( I - V G_{\rm L,R}^{\rm a,a} \right)^{-1},
\end{align}

\subsection{Recursive Green's function}
In this section, we give the recursive Green's function method to calculate the surface Green's function of the left and right materials denoted by $g^{\rm ret}_{\rm L/R}$ and $g^{\rm adv}_{\rm L/R}$ in the previous section.
By generalizing the Hamiltonian discretized in space, we represent the Hamiltonian of a finite-size system by the matrix $\mathcal{H} := \left\{ \mathcal{H}_{i,j} \right\}$ with
\begin{align}
    \mathcal{H}_{i,j}
    = \hat{h} \delta_{i,j} + \hat{t} \delta_{j,i+1} + \hat{t}^{\dag} \delta_{j,i-1},
\end{align}
where $\hat{h}$ and $\hat{t}$ are the on-site and hopping matrices, respectively.
Thus $\mathcal{H}_{j,j+1} = \hat{t}$, which is the coefficient of $\psi_{j}^{\dagger}\hat{t}\psi_{j+1}$ in the layer Hamiltonian above.
The retarded Green's function of this system is written as 
\begin{align}
    G_{\mathcal{N}}(\hat{z})^{\rm ret} = \left[ \hat{z} - \mathcal{H} \right]^{-1},
\end{align}
where $\hat{z} = \omega + i\eta$ for a simple bosonic system and $\hat{z} = (\omega + i\eta)\zeta_{\rm a}\Sigma_{z}$ for a bosonic BdG system, where these relations are determined by the commutation relation that 
\begin{align}
    [\phi_{i}, \phi_{j}^{\dag}] = I_{2} \delta_{i,j}, 
    \quad 
    [\psi_{i}, \psi_{j}^{\dag}] = \zeta_{\rm a}\Sigma_{z} \delta_{i,j},
    \quad
    \Sigma_{z} = \mathrm{diag}(\sigma_{A,A}, \sigma_{B,B}) = \mathrm{diag}(1, -1).
\end{align}
$\eta$ is a positive broadening, which we keep finite as a phenomenological magnon damping.
If we set the site index as $i = 1, 2, \cdots, N$, the surface component of the Green's function $G^{(N)\rm ret}_{N,N}$ obeys the Möbius recurrence
\begin{align}
    G^{(N)\rm ret}_{N,N} = X_{\mathcal{N}} \bullet G^{(N-1)\rm ret}_{N-1,N-1},
    \quad 
    X_{\mathcal{N}} = \mqty( \hat{0} & \hat{t}^{-1} \\ -\hat{t}^{\dag} & (\hat{z} - \hat{h})\hat{t}^{-1} ),
\end{align}
where $\bullet$ is the Möbius transformation defined as
\begin{align}
    \mqty( A & B \\ C & D ) \bullet Z := (AZ + B)(CZ + D)^{-1}.
\end{align}
Since the Möbius transformation is associative, we can obtain the surface Green's function of the left semi-infinite system as
\begin{align}
    G^{\rm ret}_{\mathcal{N},\rm L} = \lim_{N\to\infty} X_{\mathcal{N}}^{N-1} \bullet G^{(1)\rm ret}_{1,1}.
\end{align}
When we get the eigenvalues and eigenvectors of the Möbius transformation $X_{\mathcal{N}}$ as $X_{\mathcal{N}} = Q \Lambda Q^{-1}$, we order them so that $|\lambda_{1}| \le \cdots \le |\lambda_{n}| < 1 < |\lambda_{n+1}| \le \cdots \le |\lambda_{2n}|$, where $n = \dim \hat{h}$. The blocks $\hat{Q}_{12}$ and $\hat{Q}_{22}$ belong to the eigenvalues with modulus greater than one. These modes decay as the layer index runs from the surface into the left bulk, and the surface Green's function is
\begin{align}
    G^{\rm ret}_{\mathcal{N},\rm L} = Q \bullet \hat{0} = \hat{Q}_{12} \hat{Q}_{22}^{-1},
    \quad 
    Q = \mqty( \hat{Q}_{11} & \hat{Q}_{12} \\ \hat{Q}_{21} & \hat{Q}_{22} ).
\end{align}
The surface Green's function of the right semi-infinite system is obtained by the substitution $\hat{t} \to \hat{t}^{\dag}$ in this construction, again retaining the eigenvalues of modulus greater than one. The substitution exchanges the two directed hoppings and selects the modes that decay into the right bulk.

Now, we have reached the surface Green's function of the left and right materials standing alone, namely $G^{\rm ret}_{\mathcal{N},\rm L}$ and $G^{\rm ret}_{\mathcal{N},\rm R}$, respectively.
In the calculation in our paper, we need to consider the effect of the potential term of the interaction Hamiltonian, which is given by
\begin{align}
    H_{\rm I}^{\rm pot} = -\sum_{\langle i,j \rangle} \mathcal{J}_{i,j} \left( S_{z,i}^{\rm L} S_{z,j}^{\rm R} 
    \right)
    =
    \begin{cases}
        -\zeta_{\rm AM} \sum_{\langle i,j \rangle} \mathcal{J}_{i,j} \left( S - n^{\rm L}_{i} \right) \sigma_{l_{j}} \left( S - n^{\rm R}_{j} \right) & ({\rm FM/AM})
        \\
        -\zeta_{\rm AM} \sum_{\langle i,j \rangle} \mathcal{J}_{i,j} \sigma_{l_{i}} \left( S - n^{\rm L}_{i} \right) \sigma_{l_{j}} \left( S - n^{\rm R}_{j} \right) & ({\rm AFM/AM})
    \end{cases} 
\end{align}
By neglecting the constant term and the higher-order terms in the bosonic operators, the same Hamiltonian gives
\begin{align}
    H_{\rm I}^{\rm pot} \simeq
    \begin{cases}
        \zeta_{\rm AM} S \sum_{\langle i,j \rangle} \mathcal{J}_{i,j} \left( n^{\rm L}_{i} + n^{\rm R}_{j} \right) \sigma_{l_{j}} & ({\rm FM/AM})
        \\
        \zeta_{\rm AM} S \sum_{\langle i,j \rangle} \mathcal{J}_{i,j} \sigma_{l_{i}} \left( n^{\rm L}_{i} + n^{\rm R}_{j} \right) \sigma_{l_{j}} & ({\rm AFM/AM}),
    \end{cases}
\end{align}
where the first and second lines correspond to the ferromagnetic and antiferromagnetic/altermagnetic cases, respectively, $l_{i}$ denotes the sublattice of the site $i$, $n^{\rm L/R}_{i}$ is the number operator of the canonical magnon, and $\zeta_{\rm AM}$ enters through $S_{z}$ of the altermagnet while $\zeta_{\rm L} = 1$.
As the simplest example, we impose that the interface interaction takes a finite value only for the nearest-neighbor same-sublattice sites.
Thus, the Hamiltonian matrix of the potential term is given by
\begin{align}
    \hat{J}_{\rm L}
    =
    \begin{cases}
        \zeta_{\rm AM} S_{\rm R} \mqty(
            \mathcal{J}_{\rm AA} & 0 \\
            0 & -\mathcal{J}_{\rm BB}
        ) & ({\rm FM/AM})
        \\
        \zeta_{\rm AM} S_{\rm R} \mqty(
            \mathcal{J}_{\rm AA} & 0 \\
            0 & \mathcal{J}_{\rm BB}
        ) & ({\rm AFM/AM})
    \end{cases} ,
    \quad 
    \hat{J}_{\rm R}
    =
    \begin{cases}
        \zeta_{\rm AM} S_{\rm L} \mqty(
            \mathcal{J}_{\rm AA} & 0 \\
            0 & -\mathcal{J}_{\rm BB}
        ) & ({\rm FM/AM})
        \\
        \zeta_{\rm AM} S_{\rm L} \mqty(
            \mathcal{J}_{\rm AA} & 0 \\
            0 & \mathcal{J}_{\rm BB}
        ) & ({\rm AFM/AM})
    \end{cases}.
\end{align}
Using this potential term, the surface Green's functions of the left and right materials are modified as
\begin{align}
    g^{\rm ret}_{\rm L} = \left[ \hat{z} - \hat{h}_{\rm L} - \hat{J}_{\rm L} - \hat{t}_{\rm L}^{\dag} G^{\rm ret}_{\rm L} \hat{t}_{\rm L} \right]^{-1},
    \quad
    g^{\rm ret}_{\rm R} = \left[ \hat{z} - \hat{h}_{\rm R} - \hat{J}_{\rm R} - \hat{t}_{\rm R} G^{\rm ret}_{\rm R} \hat{t}_{\rm R}^{\dag} \right]^{-1}.
\end{align}

\section{Sublattice selection rule and the diagonal approximation}
This section gives the lowest-order expression of the spin current behind the sublattice selection rule of the main text, the diagonal approximation that neglects the inter-sublattice cross term, its test against the dependence on the interface asymmetry $r = J^{\rm ex}_{\rm BB}/J^{\rm ex}_{\rm AA}$, and the dependence of the residual current of the blocked domain on the damping and on the inter-sublattice exchange.

\subsection{Angle dependence for the asymmetric interface coupling}
\begin{figure}
    \centering 
    \includegraphics[scale=0.8]{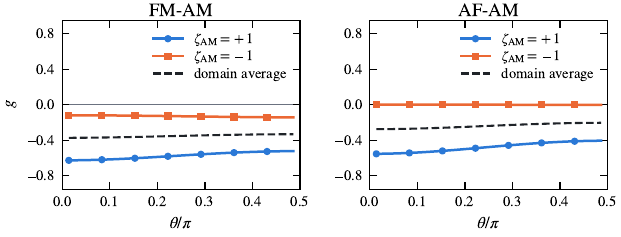}
    \caption{The dimensionless spin current $g$ as a function of the interface angle $\theta$ for the asymmetric interface coupling $J^{\rm ex}_{\rm AA} = -0.1J$ and $J^{\rm ex}_{\rm BB} = -0.05J$, i.e. $r = J^{\rm ex}_{\rm BB}/J^{\rm ex}_{\rm AA} = 0.5$.
    The panels, curves, and the other parameters are the same as in Fig.~2 of the main text.
    Here, $g$ is normalized by $((J^{\rm ex}_{\rm AA})^{2} + (J^{\rm ex}_{\rm BB})^{2})/2$ according to the definition of $g$ in the main text.}
    \label{fig:SM-g-asym}
\end{figure}

Figure~\ref{fig:SM-g-asym} shows the interface angle dependence of the normalized spin current for the asymmetric coupling $r = J^{\rm ex}_{\rm BB}/J^{\rm ex}_{\rm AA} = 0.5$, which is the counterpart of Fig.~2 of the main text.
The normalization by $((J^{\rm ex}_{\rm AA})^{2} + (J^{\rm ex}_{\rm BB})^{2})/2$ removes the trivial reduction of the overall scale.
Both characteristic features of the symmetric case are lost.
For the FM/AM junction, the two magnetic domains are no longer mirror images of each other about $\theta = \pi/4$; the current for $\zeta_{\rm AM} = +1$ is much larger than that for $\zeta_{\rm AM} = -1$, and the domain-averaged current acquires a finite $\cos 2\theta$ dependence.
For the AFM/AM junction, the current for $\zeta_{\rm AM} = +1$ does not change its sign at $\theta = \pi/4$ but has a finite uniform component over the whole range of $\theta$, while the current for $\zeta_{\rm AM} = -1$ almost vanishes.
The selection rule and the diagonal approximation below account for these features.

\subsection{Harmonic decomposition}
With $J^{\rm ex}_{\rm AB} = J^{\rm ex}_{\rm BA} = 0$, the interface angle $\theta$ enters the junction only through the altermagnetic term, whose long-wavelength form contains $\cos 2\theta$ and $\sin 2\theta$, and the current is even under $\theta \to -\theta$ and $\pi$-periodic, as shown in the main text.
The current is therefore a series in $\cos 2n\theta$,
\begin{align}
    j_{S_{z}}(\theta) = c_{0} + c_{2} \cos 2\theta + c_{4} \cos 4\theta + c_{6} \cos 6\theta + \cdots .
\end{align}
At $|J_{0}| = 0.1J$, the least-squares fit of $\sum_{n \leq 3} c_{2n} \cos 2n\theta$ to the angle sweep of Fig.~2 of the main text ($N_{\theta} = 71$ angles in $0 < \theta < \pi/2$) gives $c_{4}/c_{2} = 2.5\%$ and $c_{6}/c_{2} = 2.1 \times 10^{-4}$ for the FM/AM junction, and $c_{6}/c_{2} = 2.6 \times 10^{-4}$ ($1.0 \times 10^{-3}$) for the transmitting (blocked) domain of the AFM/AM junction, in which $c_{4}$ is forbidden by symmetry; we therefore keep $c_{0}$ and $c_{2}$ only.
The $c_{4}$ of the FM/AM junction is even under reversal of $\zeta_{\rm AM}$, like $c_{0}$, so that the domain average of the FM/AM current retains, besides $c_{0}$, a $\cos 4\theta$ term of $0.22\%$ of $c_{0}$; the cancellation of the anisotropy on the domain average stated in the main text refers to the $\cos 2\theta$ component.
The three-angle extraction below is used for the $r$ sweep of Fig.~\ref{fig:SM-r-sweep}, and the five-angle fit including $c_{4}$ for the $|J_{0}|$ sweep of Fig.~3 of the main text.
The components shown in Fig.~\ref{fig:SM-r-sweep} are extracted from the currents at the three angles $\theta/\pi = 0.1, 0.25, 0.4$ as
\begin{align}
    c_{2} = \frac{j_{S_{z}}(0.1\pi) - j_{S_{z}}(0.4\pi)}{2\cos 0.2\pi},
    \quad
    c_{0} = \frac{E + j_{S_{z}}(\pi/4) \cos 0.4\pi}{1 + \cos 0.4\pi},
    \quad
    E = \frac{1}{2} \left[ j_{S_{z}}(0.1\pi) + j_{S_{z}}(0.4\pi) \right],
\end{align}
where the $\cos 6\theta$ term is neglected.

\subsection{Lowest-order current and cross terms}
To the second order in $V$, the Keldysh component reduces to $G^{\rm K}_{\rm LR} \simeq F_{\rm L,R} = g^{\rm ret}_{\rm L} V g^{\rm K}_{\rm R} + g^{\rm K}_{\rm L} V g^{\rm adv}_{\rm R}$.
The fluctuation-dissipation theorem gives $g^{\rm K}_{\rm a} = -i \mathcal{A}_{\rm a} [1 + 2n(\omega; T_{\rm a})]$ with the Hermitian surface spectral matrix $\mathcal{A}_{\rm a} = i(g^{\rm ret}_{\rm a} - g^{\rm adv}_{\rm a})$ of the magnet $\mathrm{a}$, where $g^{\rm ret}_{\rm a} = [(\omega + i\eta)\zeta_{\rm a}\Sigma_{z} - h_{\rm a}]^{-1}$ already contains the metric of the boundary spinor; no further factor of $\Sigma_{z}$ appears, since the theorem holds componentwise for any pair of bosonic operators, and the hole-like components are accounted for by the negative spectral weight at $\omega < 0$ together with $n(-\omega; T) = -1 - n(\omega; T)$.
Writing $g^{\rm ret}_{\rm a} = \bar{g}_{\rm a} - i\mathcal{A}_{\rm a}/2$ and $g^{\rm adv}_{\rm a} = \bar{g}_{\rm a} + i\mathcal{A}_{\rm a}/2$ with the Hermitian part $\bar{g}_{\rm a}$, the terms containing $\bar{g}_{\rm a}$ are $i$ times the trace of a product of Hermitian matrices and drop out of the real part, and the remaining terms combine into $\frac{1}{2} \mathrm{Tr}_{\rm R}[V_{\rm RL} \mathcal{A}_{\rm L} V_{\rm LR} \mathcal{A}_{\rm R}] \{ [1 + 2n(\omega; T_{\rm L})] - [1 + 2n(\omega; T_{\rm R})] \}$.
The current of the main text, $j_{S_{z}} = -\mathrm{Re}\,\mathrm{Tr}_{\rm R}[V_{\rm RL} G^{\rm K}_{\rm LR}]$, therefore becomes
\begin{align}
    j_{S_{z}}^{(2)}
    = - \int_{-k_{\rm c}}^{k_{\rm c}} \frac{dk}{2\pi} \int_{-\infty}^{\infty} \frac{d\omega}{2\pi}
    \left[ n(\omega; T_{\rm L}) - n(\omega; T_{\rm R}) \right]
    \sum_{l, m \in \{ {\rm A}, {\rm B} \}} J^{\rm ex}_{ll} J^{\rm ex}_{mm}
    \left( \mathcal{A}_{\rm L} \right)_{lm} \left( \mathcal{A}_{\rm R} \right)_{ml},
    \label{eq:SM-lowest-order}
\end{align}
where the spectral matrices are taken in the boundary-spinor basis of the main text, and the diagonal hopping $V_{\rm LR} = \mathrm{diag}(J^{\rm ex}_{\rm AA}, J^{\rm ex}_{\rm BB})$ has been used.
The diagonal terms $l = m$ are the contributions of the two sublattice pairs (A, A) and (B, B), $(J^{\rm ex}_{\rm AA})^{2} j_{\rm A} + (J^{\rm ex}_{\rm BB})^{2} j_{\rm B}$.
The off-diagonal terms $l \neq m$ are the cross term $2 J^{\rm ex}_{\rm AA} J^{\rm ex}_{\rm BB} j_{\rm AB}$, which involves the inter-sublattice elements of the spectral matrices: the normal inter-sublattice element in the ferromagnet, and the anomalous element in the antiferromagnet and the altermagnet, which is of the order of the Bogoliubov admixture $u_{\vec{k}} v_{\vec{k}}$ of the two sublattices in the magnon eigenmodes and vanishes for $J_{0} \to 0$.
The selection rule of the main text concerns the diagonal terms.
For a pair of parallel spins, the product $(\mathcal{A}_{\rm L})_{ll} (\mathcal{A}_{\rm R})_{ll}$ connects the particle-like spectral weights of the two magnets, whereas for a pair of antiparallel spins it connects a particle-like weight with a hole-like one and is of the order of $(u_{\vec{k}} v_{\vec{k}})^{2}$.
The residual current of a blocked pair is therefore of the same order $V^{2}$ as the current of a transmitting pair, suppressed by the admixture rather than by the damping $\eta$.

We separate the cross term numerically, without approximation, by the additivity defect
\begin{align}
    D(\lambda) = j_{S_{z}}(\lambda J^{\rm ex}_{\rm AA}, \lambda J^{\rm ex}_{\rm BB}) - j_{S_{z}}(\lambda J^{\rm ex}_{\rm AA}, 0) - j_{S_{z}}(0, \lambda J^{\rm ex}_{\rm BB}),
    \label{eq:SM-additivity}
\end{align}
computed with the full Dyson resummation for $\lambda = 0.25, 0.5, 1$: the coefficient of $\lambda^{2}$ in $D(\lambda)$ is the cross term, and the remainder collects the higher orders in $J^{\rm ex}$ that do not add over the pairs.
The components with a single pair, $j_{S_{z}}(\lambda J^{\rm ex}_{\rm AA}, 0)$ and $j_{S_{z}}(0, \lambda J^{\rm ex}_{\rm BB})$, are related by the combination of the spin rotation $S_{z} \to -S_{z}$, the sublattice exchange, and the reflection $\theta \to \pi/2 - \theta$ of the main text, $j_{S_{z}}(J^{\rm ex}_{\rm AA}, J^{\rm ex}_{\rm BB}; \theta) = -j_{S_{z}}(J^{\rm ex}_{\rm BB}, J^{\rm ex}_{\rm AA}; \pi/2 - \theta)$ for $\zeta_{\rm AM} = +1$ of the AFM/AM junction, which holds in the numerical results to machine precision.
Table~\ref{tab:SM-cross} lists the decomposition for the parameters of Fig.~2 of the main text.
In the FM/AM junction the cross term is $5.1\%$ of $c_{0}$ and $4.6\%$ of $c_{2}$, with the sign opposite to the pair terms, even under reversal of $\zeta_{\rm AM}$ for $c_{0}$ and odd under that reversal for $c_{2}$; it is the origin of the slight $r$ dependence of the $\zeta_{\rm AM} = +1$ components in Fig.~\ref{fig:SM-r-sweep}(a) and (b), which is linear in $r$.
In the transmitting domain of the AFM/AM junction the cross term is $1.5\%$ of $c_{2}^{+}$, smaller than in the FM/AM junction because both inter-sublattice elements are anomalous.
In the blocked domain, the residual $c_{2}^{-}$, $2.3\%$ of $c_{2}^{+}$, consists of the two blocked pair terms ($52\%$), the cross term ($67\%$), and the higher orders ($-19\%$).
The cross term of the AFM/AM junction is odd in $\zeta_{\rm AM}$ to $0.1\%$ in the numerical results, so that it drops out of the domain average $\bar{c}_{2}$ at this order.
The higher orders in $D(\lambda)$ are below $0.7\%$ of the transmitting components $c_{0}$ and $c_{2}^{+}$ in all cases (they reach $-19\%$ of the small residual $c_{2}^{-}$ of the blocked domain, i.e.\ $0.4\%$ of $c_{2}^{+}$), whereas the odd correction to the quadratic scaling of a single transmitting pair is $11\%$, which is the correction quoted in the main text; the latter adds over the pairs and does not enter $D(\lambda)$.

\begin{table}
    \caption{Decomposition of the harmonic components at $|J_{0}| = 0.1J$ and $r = 1$ (the parameters of Fig.~2 of the main text) into the sublattice-pair terms, the cross term, and the higher orders, obtained from the additivity defect, Eq.~(\ref{eq:SM-additivity}).
    The full components are given in the units of $g$ of the main text; the cross term and the higher orders are given in percent of the full component.
    The uniform component of the AFM/AM junction vanishes by symmetry.}
    \label{tab:SM-cross}
    \begin{ruledtabular}
    \begin{tabular}{llrrr}
        junction & component & full ($g$) & cross term (\%) & higher orders (\%) \\
        \hline
        FM/AM, $\zeta_{\rm AM} = \pm 1$ & $c_{0}$ & $-0.351$ & $5.1$ & $0.3$ \\
        FM/AM, $\zeta_{\rm AM} = \pm 1$ & $c_{2}$ & $\mp 0.0318$ & $4.6$ & $0.5$ \\
        AFM/AM, $\zeta_{\rm AM} = +1$ & $c_{2}$ & $-0.0753$ & $1.5$ & $0.6$ \\
        AFM/AM, $\zeta_{\rm AM} = -1$ & $c_{2}$ & $+0.00173$ & $67$ & $-19$ \\
    \end{tabular}
    \end{ruledtabular}
\end{table}

\subsection{Diagonal approximation}
\begin{figure}
    \centering
    \includegraphics[width=0.9\linewidth]{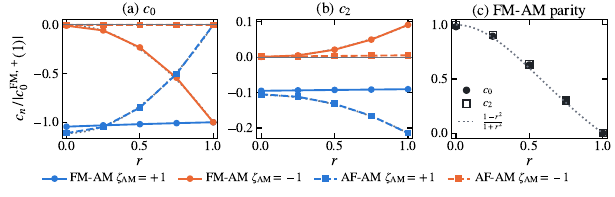}
    \caption{Test of the diagonal approximation against the dependence of the harmonic components of the spin current on the interface asymmetry $r = J^{\rm ex}_{\rm BB}/J^{\rm ex}_{\rm AA}$ with $J^{\rm ex}_{\rm AA} = -0.1J$ fixed.
    (a) The uniform component $c_{0}$ and (b) the $\cos 2\theta$ component $c_{2}$ of $j_{S_{z}}(\theta) = c_{0} + c_{2}\cos 2\theta + \cdots$, in units of the magnitude of the uniform component of the FM/AM junction with $\zeta_{\rm AM} = +1$ at $r = 1$.
    The dotted lines are proportional to $r^{2}$ and $1 - r^{2}$ in (a), and to $r^{2}$ and $(1 + r^{2})/2$ in (b), as predicted by Eqs.~(\ref{eq:SM-FMAM-channel}) and (\ref{eq:SM-AFMAM-channel}).
    (c) The domain-odd/even ratio $(c_{0}^{+} - c_{0}^{-})/(c_{0}^{+} + c_{0}^{-})$ (circles) and the domain-even/odd ratio $(c_{2}^{+} + c_{2}^{-})/(c_{2}^{+} - c_{2}^{-})$ (squares) of the FM/AM junction, where $c_{n}^{\pm}$ denotes $c_{n}$ for $\zeta_{\rm AM} = \pm 1$.
    The dotted line is $(1 - r^{2})/(1 + r^{2})$ without any fitting parameter.
    The slight $r$ dependence of the $\zeta_{\rm AM} = +1$ components in (a) and (b) is the cross term of Table~\ref{tab:SM-cross}, which is linear in $r$.
    The other parameters are the same as in Fig.~2 of the main text.}
    \label{fig:SM-r-sweep}
\end{figure}

Neglecting the cross term and the higher orders in Eq.~(\ref{eq:SM-lowest-order}), the current is the sum of the two sublattice pairs,
\begin{align}
    j_{S_{z}} \simeq (J^{\rm ex}_{\rm AA})^{2} j_{\rm A} + (J^{\rm ex}_{\rm BB})^{2} j_{\rm B},
    \label{eq:SM-diagonal}
\end{align}
where $j_{l}$ is determined by the surface spectral functions of the two electrodes on the sublattice $l$.
We call Eq.~(\ref{eq:SM-diagonal}) the diagonal approximation; at $|J_{0}| = 0.1J$ it reproduces $c_{0}$ and $c_{2}$ to about $5\%$ (Table~\ref{tab:SM-cross}), and its predictions for the dependence on the interface asymmetry $r$ are tested in Fig.~\ref{fig:SM-r-sweep}.

Whether a pair is transmitting or blocked is decided by the relative direction of the two spins it connects.
When the spins on both sides are parallel, the transverse coupling $S^{\rm L}_{+} S^{\rm R}_{-} + {\rm h.c.}$ becomes a hopping term $\phi^{\dag}\psi + {\rm h.c.}$ of a magnon across the interface, which conserves $S_{z}$ and carries the steady spin current driven by the temperature difference.
When the spins are antiparallel, the same coupling becomes a pair creation/annihilation term $\phi^{\dag}\psi^{\dag} + {\rm h.c.}$ of magnons on both sides.
Such a term transmits a magnon only through the Bogoliubov admixture of the two sublattices in the eigenmodes, so that the pair carries a residual current of the same order $V^{2}$ but suppressed by the admixture, as discussed above.

\subsection{FM/AM junction in the diagonal approximation}
Within the diagonal approximation, the current of the FM/AM junction follows from the selection rule alone.
For $\zeta_{\rm AM} = +1$ the A sublattice of the altermagnet is parallel and the B sublattice is antiparallel to the magnetization of the ferromagnet, and vice versa for $\zeta_{\rm AM} = -1$.
Only one pair is transmitting for each domain, and the current reads
\begin{align}
    j_{S_{z}}^{\rm FM-AM}(\theta; \zeta_{\rm AM}) = (J^{\rm ex}_{\zeta_{\rm AM}})^{2} \left( a + \zeta_{\rm AM} b \cos 2\theta \right),
    \quad
    J^{\rm ex}_{+1} = J^{\rm ex}_{\rm AA}, \quad J^{\rm ex}_{-1} = J^{\rm ex}_{\rm BB},
    \label{eq:SM-FMAM-channel}
\end{align}
where $a$ and $b$ are independent of $\zeta_{\rm AM}$ and $r$.
The sign $\zeta_{\rm AM}$ in front of $b$ follows from the fact that the domain reversal $\zeta_{\rm AM} \to -\zeta_{\rm AM}$ is equivalent to the simultaneous sublattice exchange ${\rm A} \leftrightarrow {\rm B}$ and the reversal $\delta J \to -\delta J$, while the ferromagnet cannot distinguish the two sublattices.
Equation~(\ref{eq:SM-FMAM-channel}) gives $c_{0}^{\pm} = (J^{\rm ex}_{\pm 1})^{2} a$ and $c_{2}^{\pm} = \pm (J^{\rm ex}_{\pm 1})^{2} b$, hence $c_{0}^{+}$ and $c_{2}^{+}$ are independent of $r$, $c_{0}^{-}$ and $c_{2}^{-}$ are proportional to $r^{2}$, and
\begin{align}
    \frac{c_{0}^{+} - c_{0}^{-}}{c_{0}^{+} + c_{0}^{-}}
    =
    \frac{c_{2}^{+} + c_{2}^{-}}{c_{2}^{+} - c_{2}^{-}}
    =
    \frac{1 - r^{2}}{1 + r^{2}},
\end{align}
which is the dotted line in Fig.~\ref{fig:SM-r-sweep}(c).
At $r = 1$ the uniform component is even under reversal of $\zeta_{\rm AM}$ and the $\cos 2\theta$ component is odd under that reversal, which reproduces the form $j_{S_{z}}(\theta; \zeta_{\rm AM}) = c_{0} + \zeta_{\rm AM} c_{2}^{+} \cos 2\theta$ of Fig.~2 of the main text, so that the $\cos 2\theta$ component cancels on the domain average $\bar{g}$.
For $r \neq 1$ the domain-even part of $c_{2}$, proportional to $(J^{\rm ex}_{\rm AA})^{2} - (J^{\rm ex}_{\rm BB})^{2}$, survives the domain average.

\subsection{AFM/AM junction in the diagonal approximation}
For $\zeta_{\rm AM} = +1$ the spins on both sublattices of the altermagnet are parallel to those of the antiferromagnet, so that both pairs are transmitting; for $\zeta_{\rm AM} = -1$ both are antiparallel and both pairs are blocked, which is why the $\zeta_{\rm AM} = -1$ current is strongly suppressed in Fig.~2 of the main text, Fig.~\ref{fig:SM-r-sweep}, and Fig.~\ref{fig:SM-g-asym}.
For $\zeta_{\rm AM} = +1$, the magnon on the A sublattice carries $S_{z} = -1$ and that on the B sublattice carries $S_{z} = +1$.
The uniform currents of the two pairs thus have the same magnitude and opposite signs, $j_{\rm B}^{(0)} = -j_{\rm A}^{(0)}$.
For the $\cos 2\theta$ component, the sublattice exchange reverses $\delta J \cos 2\theta$ as well, and the two sign changes compensate each other, $j_{\rm B}^{(2)} = +j_{\rm A}^{(2)}$.
Therefore,
\begin{align}
    j_{S_{z}}^{\rm AFM-AM}(\theta; +1) = \left( (J^{\rm ex}_{\rm AA})^{2} - (J^{\rm ex}_{\rm BB})^{2} \right) a' + \left( (J^{\rm ex}_{\rm AA})^{2} + (J^{\rm ex}_{\rm BB})^{2} \right) b' \cos 2\theta,
    \label{eq:SM-AFMAM-channel}
\end{align}
i.e. $c_{0}^{+} \propto 1 - r^{2}$ and $c_{2}^{+} \propto (1 + r^{2})/2$, which are the dotted lines in Fig.~\ref{fig:SM-r-sweep}(a) and (b).
The vanishing uniform component at $r = 1$ is the realization of the symmetry argument of the main text within the diagonal approximation, and the uniform component at $r \neq 1$ is a direct measure of the asymmetry of the interface coupling.

\subsection{Residual current of the blocked domain}
\begin{figure}
    \centering
    \includegraphics[scale = 1.0]{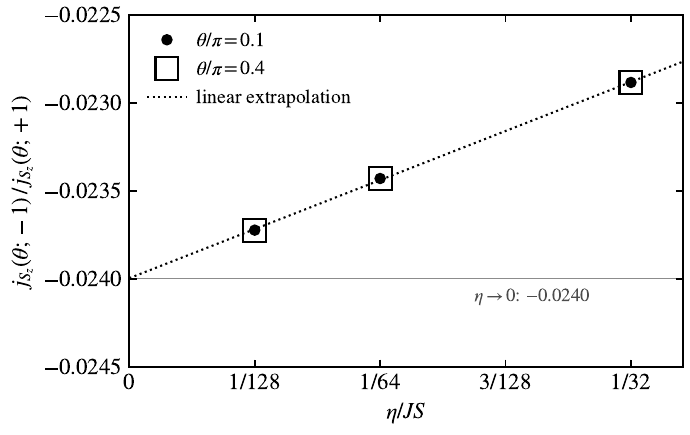}
    \caption{Ratio of the currents of the two domains of the AFM/AM junction, $j_{S_{z}}(\theta; \zeta_{\rm AM} = -1)/j_{S_{z}}(\theta; \zeta_{\rm AM} = +1)$, as a function of the damping $\eta$ at $\theta/\pi = 0.1$ (circles) and $0.4$ (squares).
    The two angles coincide to machine precision.
    The dotted line is the linear fit, and the thin horizontal line marks its extrapolation $-0.0240$ at $\eta \to 0$.
    The other parameters are the same as in Fig.~2 of the main text.}
    \label{fig:SM-eta}
\end{figure}

Figure~\ref{fig:SM-eta} shows the ratio of the currents of the two domains of the AFM/AM junction, $j_{S_{z}}(\theta; \zeta_{\rm AM} = -1)/j_{S_{z}}(\theta; \zeta_{\rm AM} = +1)$, for $\eta = JS/32$, $JS/64$, and $JS/128$ at $\theta/\pi = 0.1$ and $0.4$.
The ratio is $-0.0229$, $-0.0234$, and $-0.0237$ for $\eta = JS/32$, $JS/64$, and $JS/128$, independent of $\theta$ to machine precision as required by the pure $\cos 2\theta$ form of both domains, and extrapolates linearly to $-0.0240$ at $\eta \to 0$; its magnitude slightly increases with decreasing $\eta$.
The residual current is thus not an artifact of the damping but the transmission through the admixture of the sublattices, in accordance with Eq.~(\ref{eq:SM-lowest-order}).

The dependence on the inter-sublattice exchange in Fig.~3 of the main text is computed at the fixed $\overline{T} = JS$, $K = 0.05J$, and $\eta = JS/32$, so that the magnon gap of the antiferromagnet and the altermagnet, $\Delta = \sqrt{16K|J_{0}| + 4K^{2}}\,S$, increases from $0.3JS$ at $|J_{0}| = 0.1J$ to $0.9JS$ at $|J_{0}| = J$.
The harmonic components are extracted by the least-squares fit of $c_{0}$, $c_{2}$, and $c_{4}$ to the currents at the five angles $\theta/\pi = 0.05, 0.1, 0.25, 0.4, 0.45$.
The ratio $c_{2}^{-}/c_{2}^{+}$ changes from $-0.023$ at $|J_{0}| = 0.1J$ to $-0.25$ at $|J_{0}| = J$; its magnitude grows almost linearly with $|J_{0}|$, and the domain average $\bar{c}_{2}$ retains $49\%$ to $38\%$ of $c_{2}^{+}$.
The uniform component of the AFM/AM junction and the domain-averaged $\cos 2\theta$ component of the FM/AM junction vanish by symmetry for all $|J_{0}|$.
The current of the AFM/AM junction remains a pure $\cos 2\theta$ up to $|J_{0}| = J$, with $c_{4}$ forbidden by the reflection $j_{S_{z}}(\theta) = -j_{S_{z}}(\pi/2 - \theta)$, whereas the FM/AM junction develops $c_{4}/c_{2} \simeq 0.05$ at $|J_{0}| = J$.

\subsection{Interface registry and the sign of the interface exchange}
The results of the main text are computed for the same-sublattice registry, $\mathcal{J}_{\rm AA} = \mathcal{J}_{\rm BB}$ and $\mathcal{J}_{\rm AB} = \mathcal{J}_{\rm BA} = 0$, in which the bonded sites have no in-plane displacement.
The cross-sublattice registry, $\mathcal{J}_{\rm AB} = \mathcal{J}_{\rm BA}$ and $\mathcal{J}_{\rm AA} = \mathcal{J}_{\rm BB} = 0$, bonds the site A of one magnet to the site B of the other, displaced by $(a/2, a/2)$, whose in-plane component $\frac{a}{2}(\cos\theta - \sin\theta)$ enters the hopping as the phase $e^{ik_{\parallel} a (\cos\theta - \sin\theta)/2}$.
The relabeling $\mathrm{A} \leftrightarrow \mathrm{B}$ of the altermagnet maps $(\zeta_{\rm AM}, \delta J)$ onto $(-\zeta_{\rm AM}, -\delta J)$ and the same-sublattice coupling onto the cross-sublattice one at fixed bond geometry; combined with the reflection relation $j_{S_{z}}(\theta; \delta J) = j_{S_{z}}(\pi/2 - \theta; -\delta J)$ of the main text, it gives $j_{S_{z}}^{\rm AB}(\theta; \zeta_{\rm AM}) = -j_{S_{z}}^{\rm AA}(\theta; -\zeta_{\rm AM})$ for the AFM/AM junction and $j_{S_{z}}^{\rm AB}(\theta; \zeta_{\rm AM}) = j_{S_{z}}^{\rm AA}(\theta; \zeta_{\rm AM})$ for the FM/AM junction, exactly when the displacement phase vanishes.
Table~\ref{tab:SM-registry} compares the two registries at the parameters of Fig.~2 of the main text: the relations hold to machine precision at $\theta = \pi/4$, where the phase vanishes, and to $0.03$--$0.24\%$ for the transmitting domain and the FM/AM junction, and to $1.3\%$ for the blocked domain, at the other angles; the residual is smooth in $\theta$ and is the effect of the displacement.
For the same reason, the symmetry relations $c_{0} = 0$ (AFM/AM) and $c_{2}^{+} = -c_{2}^{-}$ (FM/AM) hold exactly for the same-sublattice registry and to $10^{-4}$--$6 \times 10^{-3}$ for registries with displaced bonds.
The transmitting domain of the cross-sublattice registry is $\zeta_{\rm AM} = -1$, with $c_{2}$ of the opposite sign: the selection rule concerns the spins connected by the bonds.

\begin{table}
    \caption{Harmonic components of the same-sublattice registry (AA, $J^{\rm ex}_{\rm AA} = J^{\rm ex}_{\rm BB} = -0.1J$) and of the cross-sublattice registry (AB, $J^{\rm ex}_{\rm AB} = J^{\rm ex}_{\rm BA} = -0.1J$), of the mixed registries with $J^{\rm ex}_{\rm AA} = J^{\rm ex}_{\rm BB} = J_{d}$ and $J^{\rm ex}_{\rm AB} = J^{\rm ex}_{\rm BA} = J_{o}$ at the fixed total weight $2(J_{d}^{2} + J_{o}^{2}) = 0.02J^{2}$, and of the same-sublattice registry with an antiferromagnetic interface exchange ($J^{\rm ex}_{\rm AA} = J^{\rm ex}_{\rm BB} = +0.1J$), at the parameters of Fig.~2 of the main text.
    All components are given in the units of $g$ of the main text with the common normalization $2J^{2}/(0.02J^{2})$; $c_{0}$ of the AFM/AM junction is below $10^{-4}$ of $c_{2}$ in all rows.
    The last column is the domain average $\bar{c}_{2}$ of the AFM/AM junction in units of its value for the AA registry, compared with $(J_{d}^{2} - J_{o}^{2})/(J_{d}^{2} + J_{o}^{2})$ in parentheses.}
    \label{tab:SM-registry}
    \begin{ruledtabular}
    \begin{tabular}{llrrrrr}
        registry & $(J_{d}, J_{o})/0.1J$ & FM/AM $c_{0}$ & FM/AM $c_{2}^{+}$ & AFM/AM $c_{2}^{+}$ & AFM/AM $c_{2}^{-}$ & $\bar{c}_{2}/\bar{c}_{2}^{\rm AA}$ \\
        \hline
        AA & $(1, 0)$ & $-0.351$ & $-0.0318$ & $-0.0753$ & $0.00172$ & $1$ ($1$) \\
        AB & $(0, 1)$ & $-0.351$ & $-0.0318$ & $-0.00170$ & $0.0753$ & $-1.00$ ($-1$) \\
        mixed & $(1, 1)/\sqrt{2}$ & $-0.335$ & $-0.0339$ & $-0.0238$ & $0.0238$ & $0$ ($0$) \\
        mixed & $(1, 2)/\sqrt{5}$ & $-0.340$ & $-0.0336$ & $-0.00652$ & $0.0472$ & $-0.553$ ($-0.6$) \\
        mixed & $(2, 1)/\sqrt{5}$ & $-0.340$ & $-0.0336$ & $-0.0473$ & $0.00654$ & $0.553$ ($0.6$) \\
        AA, $\mathcal{J} < 0$ & $(+1, 0)$ & $-0.280$ & $-0.0251$ & $-0.0573$ & $0.00312$ & $0.74$ \\
    \end{tabular}
    \end{ruledtabular}
\end{table}

When both couplings coexist with $\mathcal{J}_{\rm AA} = \mathcal{J}_{\rm BB}$ and $\mathcal{J}_{\rm AB} = \mathcal{J}_{\rm BA}$, the combined symmetry of the main text still holds, $c_{0}$ of the AFM/AM junction vanishes, and the two registries open opposite domains with opposite signs of $c_{2}$.
In the diagonal approximation the domain average is $\bar{c}_{2} \simeq \frac{1 - \epsilon}{2} \frac{J_{d}^{2} - J_{o}^{2}}{J_{d}^{2} + J_{o}^{2}} \, c_{2}^{\rm AA, +}$ with the leak $\epsilon = 0.023$ of the blocked domain; Table~\ref{tab:SM-registry} confirms this rule to $2.3\%$ of $c_{2}^{\rm AA, +}$, whereas the components of the individual domains fall short of the diagonal approximation by $0.1$--$0.17\, c_{2}^{\rm AA, +}$.
The additivity defect of Eq.~(\ref{eq:SM-additivity}), applied to the two registries at $J_{d} = J_{o}$, attributes this deficit to the cross term between the same-sublattice and the cross-sublattice bonds, $44\%$ of $c_{2}$ in each domain (higher orders $13\%$), much larger than the cross term between the (A, A) and (B, B) pairs of Table~\ref{tab:SM-cross} because it involves the normal inter-sublattice spectral function of the altermagnet; it has the sign opposite to $c_{2}$ in each domain and cancels in the domain average.
For equal couplings, $J_{d} = J_{o}$, the two domains carry equal and opposite $c_{2}$, the valve is lost, and the domain average vanishes; the anisotropy survives the domain average unless the two couplings are equal, and for a lattice continuous across a compensated interface, where the nearest-neighbor bonds are cross-sublattice, it is dominated by the cross-sublattice channel.

The sign of the interface exchange enters through the interface potentials $U^{\rm L/R}$ and the odd orders of the Dyson series.
For the antiferromagnetic interface exchange $\mathcal{J}_{\rm AA} = \mathcal{J}_{\rm BB} = -0.1J$ (last row of Table~\ref{tab:SM-registry}), the symmetry relations hold to machine precision, the transmitting domain is unchanged, the currents of the transmitting domain and of the FM/AM junction decrease by $20$--$24\%$, and the leak of the blocked domain increases to $c_{2}^{-}/c_{2}^{+} = -0.054$.
The sign of the interface exchange thus decides which configuration is the ground state of the junction (parallel bonded spins for $\mathcal{J} > 0$, antiparallel for $\mathcal{J} < 0$) and changes the magnitude of the current, but not the selection rule.

\subsection{Sensitivity to the cutoff and the temperature}
The long-wavelength form of the dispersion is used up to the cutoff $k_{\rm c} = 1/a$, at which the lattice dispersion $2JS[1 - \cos(ka)]$ deviates from its quadratic expansion by $8\%$ ($2\%$ at $k_{\rm c} = 0.5/a$ and $17\%$ at $1.5/a$).
The cutoff is a parameter of the model rather than a convergence parameter, and the symmetry relations hold for any $k_{\rm c}$.
Table~\ref{tab:SM-kc} lists the harmonic components at $k_{\rm c} = 0.5/a$, $1/a$, and $1.5/a$ for the parameters of Fig.~2 of the main text: the absolute scale of $g$ changes, $c_{2}/c_{0}$ of the FM/AM junction changes by at most $6\%$ and $c_{2}^{-}/c_{2}^{+}$ of the AFM/AM junction by at most $22\%$, both at the smaller cutoff ($3\%$ at $k_{\rm c} = 1.5/a$), whereas the domain-averaged ratio $\bar{c}_{2}/c_{2}^{+}$ changes by less than $1\%$.

\begin{table}
    \caption{Harmonic components for the cutoff $k_{\rm c} = 0.5/a$, $1/a$, and $1.5/a$ at $|J_{0}| = 0.1J$, $r = 1$, and $\overline{T} = JS$ (the parameters of Fig.~2 of the main text), extracted by the five-angle fit.
    $c_{0}$ and $c_{2}^{+}$ are given in the units of $g$ of the main text.}
    \label{tab:SM-kc}
    \begin{ruledtabular}
    \begin{tabular}{lrrrrr}
        $k_{\rm c} a$ & FM/AM $c_{0}$ & FM/AM $c_{2}/c_{0}$ & AFM/AM $c_{2}^{+}$ & AFM/AM $c_{2}^{-}/c_{2}^{+}$ & AFM/AM $\bar{c}_{2}/c_{2}^{+}$ \\
        \hline
        $0.5$ & $-0.478$ & $0.096$ & $-0.111$ & $-0.028$ & $0.486$ \\
        $1$   & $-0.351$ & $0.091$ & $-0.075$ & $-0.023$ & $0.489$ \\
        $1.5$ & $-0.251$ & $0.088$ & $-0.052$ & $-0.022$ & $0.489$ \\
    \end{tabular}
    \end{ruledtabular}
\end{table}

The applicability of the linear spin-wave theory at $\overline{T} = JS$ is checked by the thermal magnon number per site of the bulk, computed with the lattice spin-wave dispersion of the model on the two-dimensional Brillouin zone without the long-wavelength expansion.
At $T = JS$, $\langle n \rangle_{\rm th} = 0.12S$ in the ferromagnet and $0.12S$ in the antiferromagnet and the altermagnet at $|J_{0}| = 0.1J$ ($0.030S$ at $|J_{0}| = J$), and $0.17S$ at the hotter electrode $T = 1.2JS$; the zero-point reduction of the sublattice moment is $0.0066S$ at $|J_{0}| = 0.1J$ and $0.061S$ at $|J_{0}| = J$.
The dimensionless current depends on $\overline{T}$ only through the Bose factors.
Table~\ref{tab:SM-T} shows the harmonic ratios at $\overline{T} = 0.5JS$, $JS$, and $2JS$ with $\Delta T = 0.4\overline{T}$: $c_{2}^{-}/c_{2}^{+}$ changes from $-0.031$ at $0.5JS$ to $-0.021$ at $2JS$ for $|J_{0}| = 0.1J$ (from $-0.32$ to $-0.22$ at $|J_{0}| = J$), the blocked domain leaking more at lower temperature where the admixture near the band bottom dominates, and $\bar{c}_{2}/c_{2}^{+}$ stays within $0.48$--$0.49$ ($0.34$--$0.39$).
The row $\overline{T} = 2JS$, at which $\langle n \rangle_{\rm th} = 0.37S$, lies beyond the linear regime and is shown only for the trend.

\begin{table}
    \caption{Harmonic ratios for the mean temperature $\overline{T} = 0.5JS$, $JS$, and $2JS$ with $\Delta T = 0.4\overline{T}$, at $r = 1$, $k_{\rm c} = 1/a$, and $\eta = JS/32$, extracted by the five-angle fit.
    The last column is the thermal magnon number per site of the bulk antiferromagnet and altermagnet at $T = \overline{T}$.}
    \label{tab:SM-T}
    \begin{ruledtabular}
    \begin{tabular}{rrrrrr}
        $\overline{T}/JS$ & $|J_{0}|/J$ & FM/AM $c_{2}/c_{0}$ & AFM/AM $c_{2}^{-}/c_{2}^{+}$ & AFM/AM $\bar{c}_{2}/c_{2}^{+}$ & $\langle n \rangle_{\rm th}/S$ \\
        \hline
        $0.5$ & $0.1$ & $0.112$ & $-0.031$ & $0.485$ & $0.032$ \\
        $0.5$ & $1$   & $0.076$ & $-0.318$ & $0.341$ & $0.0050$ \\
        $1$   & $0.1$ & $0.091$ & $-0.023$ & $0.489$ & $0.117$ \\
        $1$   & $1$   & $0.082$ & $-0.251$ & $0.375$ & $0.030$ \\
        $2$   & $0.1$ & $0.076$ & $-0.021$ & $0.490$ & $0.371$ \\
        $2$   & $1$   & $0.070$ & $-0.220$ & $0.390$ & $0.127$ \\
    \end{tabular}
    \end{ruledtabular}
\end{table}

\subsection{Inter-sublattice exchange interaction dependence}

\begin{figure}
    \centering 
    \includegraphics[scale = 1.0]{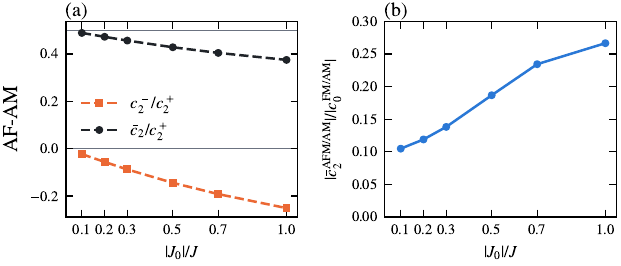}
    \caption{Dependence on the inter-sublattice exchange $|J_{0}|/J$ for the symmetric coupling $r = J^{\rm ex}_{\rm BB}/J^{\rm ex}_{\rm AA} = 1$.
    (a) The ratio of the blocked to the transmitting domain, $c_{2}^{-}/c_{2}^{+}$, and the domain-averaged $\cos 2\theta$ component $\bar{c}_{2}/c_{2}^{+}$ of the AFM/AM junction, where $\bar{c}_{n} = (c_{n}^{+} + c_{n}^{-})/2$.
    The horizontal lines mark a perfect valve, $c_{2}^{-} = 0$ and $\bar{c}_{2}/c_{2}^{+} = 1/2$.
    (b) The domain-averaged $\cos 2\theta$ component of the AFM/AM junction in units of the uniform component of the FM/AM junction.
    $J_{0}$ is ferromagnetic in the FM and antiferromagnetic in the AFM and the AM, and the other parameters are the same as in the main text, so that the magnon gap of the AFM and the AM, $\sqrt{16K|J_{0}| + 4K^{2}}\,S$, increases with $|J_{0}|$ at the fixed $\overline{T}$.}
    \label{fig:J0_sweep}
\end{figure}

Both arguments rely on the symmetric coupling, since $\mathrm{A} \leftrightarrow \mathrm{B}$ also exchanges $J^{\rm ex}_{\rm AA} \leftrightarrow J^{\rm ex}_{\rm BB}$.
For the asymmetric coupling $r = J^{\rm ex}_{\rm BB}/J^{\rm ex}_{\rm AA} \neq 1$, the two domains of the FM/AM junction are no longer mirror images about $\theta = \pi/4$, and the AFM/AM junction acquires a finite uniform component.
These features originate in the relative orientation of the spins coupled across the interface.
Since the interface couples only the same sublattices, the current to leading order in $J^{\rm ex}$ is a sum over the two sublattice pairs (A, A) and (B, B), each weighted by the product of the surface spectral functions of the two magnets on that pair, plus a cross term involving the inter-sublattice spectral functions, which amounts to a few percent of the pair terms at $|J_{0}| = 0.1J$ except in the blocked AFM/AM domain, where it dominates the residual current.
When the two spins of a pair are parallel, the transverse exchange $S^{\rm L}_{+} S^{\rm R}_{-} + {\rm h.c.}$ is a hopping $\phi^{\dagger}\psi + {\rm h.c.}$ of a magnon across the interface, which conserves $S_{z}$ and transmits the thermal magnons.
When they are antiparallel, it is a pair term $\phi^{\dagger}\psi^{\dagger} + {\rm h.c.}$, which transmits a magnon only through the admixture of the two sublattices in the magnon eigenmodes caused by the inter-sublattice exchange $J_{0}$.
In the AFM/AM junction, both pairs are parallel for $\zeta_{\rm AM} = +1$ and antiparallel for $\zeta_{\rm AM} = -1$; the current is transmitted in the former and blocked in the latter, which is the magnon valve.
In the FM/AM junction, one pair is parallel and the other antiparallel in either domain, so that the two domains transmit comparable currents and are related by the reflection about $\theta = \pi/4$.
The residual current of the blocked domain is a measure of the admixture: the ratio $c_{2}^{-}/c_{2}^{+}$ converges to $-0.024$ as $\eta \to 0$ and its magnitude grows almost linearly with $|J_{0}|/J$, as shown in Fig.~\ref{fig:J0_sweep}(a); for an antiferromagnetic interface exchange of the same magnitude, the interface potentials change the currents by about $20\%$ and the ratio becomes $-0.054$, with the same transmitting domain.
Even at $|J_{0}| = J$, the blocked domain carries only a quarter of the current of the transmitting domain, and the domain-averaged $\cos 2\theta$ component $\bar{c}_{2} = (c_{2}^{+} + c_{2}^{-})/2$ retains $38\%$ of $c_{2}^{+}$, compared with $49\%$ at $|J_{0}| = 0.1J$ and $50\%$ for a perfect valve.
Figure~\ref{fig:J0_sweep}(b) compares the domain-averaged anisotropic current of the AFM/AM junction with the uniform current of the FM/AM junction at the same interface: the former is $10$--$27\%$ of the latter and grows with $|J_{0}|$, exceeding the $\cos 2\theta$ modulation ($6$--$9\%$) of a single domain of the FM/AM junction, which cancels on the domain average.

\end{document}